\documentclass[english,aps,pra,reprint,superscriptaddress]{revtex4-1}
\usepackage{graphicx}
\usepackage{color}
\usepackage[colorlinks=true, pdfstartview=FitV, linkcolor=blue, citecolor=blue, urlcolor=blue]{hyperref}
\usepackage{amsmath}
\usepackage{amssymb}    
\usepackage{subcaption} 
\usepackage[section]{placeins}

\newcommand{\sgn}{\text{sgn}}

\begin{document}
\title{Simulating Spin Waves in Entropy Stabilized Oxides}
\author{Tom Berlijn}
\email{berlijnt@ornl.gov}
\affiliation{Center for Nanophase Materials Sciences, Oak Ridge National Laboratory, Oak Ridge, TN 37831, USA}
\author{Gonzalo Alvarez}
\affiliation{Computational Sciences \& Engineering Division, Oak Ridge National Laboratory, Oak Ridge, TN 37831, USA}
\author{David S. Parker}
\affiliation{Materials Science \& Technology Division, Oak Ridge National Laboratory, Oak Ridge, TN 37831, USA}
\author{Rapha\"el P. Hermann}
\author{Randy S. Fishman}
\affiliation{Materials Science \& Technology Division, Oak Ridge National Laboratory, Oak Ridge, TN 37831, USA}

\begin{abstract}
The entropy stabilized oxide Mg$_{0.2}$Co$_{0.2}$Ni$_{0.2}$Cu$_{0.2}$Zn$_{0.2}$O exhibits antiferromagnetic order and magnetic excitations, as revealed by recent neutron scattering experiments. This observation raises the question of the nature of spin wave excitations in such disordered systems. Here, we investigate theoretically the magnetic ground state and the spin-wave excitations using linear spin-wave theory in combination with the supercell approximation to take into account the extreme disorder in this magnetic system. We find that the experimentally observed antiferromagnetic structure can be stabilized by a rhombohedral distortion together with large second nearest neighbor interactions. Our calculations show that the spin-wave spectrum consists of a well-defined low-energy coherent spectrum in the background of an incoherent continuum that extends to higher energies. 
\end{abstract}
\maketitle

\section{introduction}
Entropy stabilization means that a single phase is stabilized by the mixing entropy gained from randomly distributing a number of elements, typically five or more,  over a single crystal lattice.~\cite{miracle_2017} In 2015 a new family was added to the class of high entropy materials with the discovery of the high entropy oxides (HEOs). By mixing equimolar amounts of MgO, CoO, NiO, CuO and ZnO Rost \textit{et al.}~\cite{rost_2015_ncomm} found that Mg$_{0.2}$Co$_{0.2}$Ni$_{0.2}$Cu$_{0.2}$Zn$_{0.2}$O (hereafter MgO-HEO) could be stabilized in a simple rock-salt structure in which the oxygen atoms occupy one of the face-centered cubic (FCC) sublattices and the cations are randomly distributed over the other FCC sublattice. The entropy stabilized nature of the phase is apparent when considering that CuO and ZnO do not form in the rock-salt structure that forces Cu and Zn in octahedral coordination. Rost \textit{et al.}~\cite{rost_2015_ncomm} demonstrated that the sample could be switched between a multiphase structure and the aforementioned single phase rock-salt structure by repeated heating and cooling , thereby proving  the reversibility of the entropy-driven transition. Furthermore, a combination of X-ray diffraction, X-ray absorption fine-structure and scanning transmission electron microscopy with energy dispersive X-ray spectroscopy, showed that the MgO-HEO sample was chemically and structurally homogeneous~\cite{rost_2015_ncomm}. The HEOs are not only interesting from a basic scientific point of view, but have also exhibit promising functional properties, such as high Li-ion storage capacity and cycling stability~\cite{sarkar_2019_advmat}, colossal dielectric constants~\cite{berardan_2016_pss}, superionic conductivity~\cite{berardan_2016_jmca} and a high ratio of elastic modulus to thermal conductivity~\cite{braun_2018_advmat}. In addition to rock-salt HEOs, HEOs with perovskite~\cite{sarkar_2018_jecs,jiang_2018_sm,sharma_2018_prbr}, fluorite~\cite{gild_2018_jecs,chen_2018_jecs} and spinel~\cite{dabrowa_2018_ml} structures have been discovered, together with high entropy carbides~\cite{castle_2018_srep,yan_2018_jacs}, borides~\cite{gild_2016_srep} and chalcogenides~\cite{deng_2020_chemmat}. 

\begin{figure}[!htb]
\includegraphics[width=0.8\columnwidth]{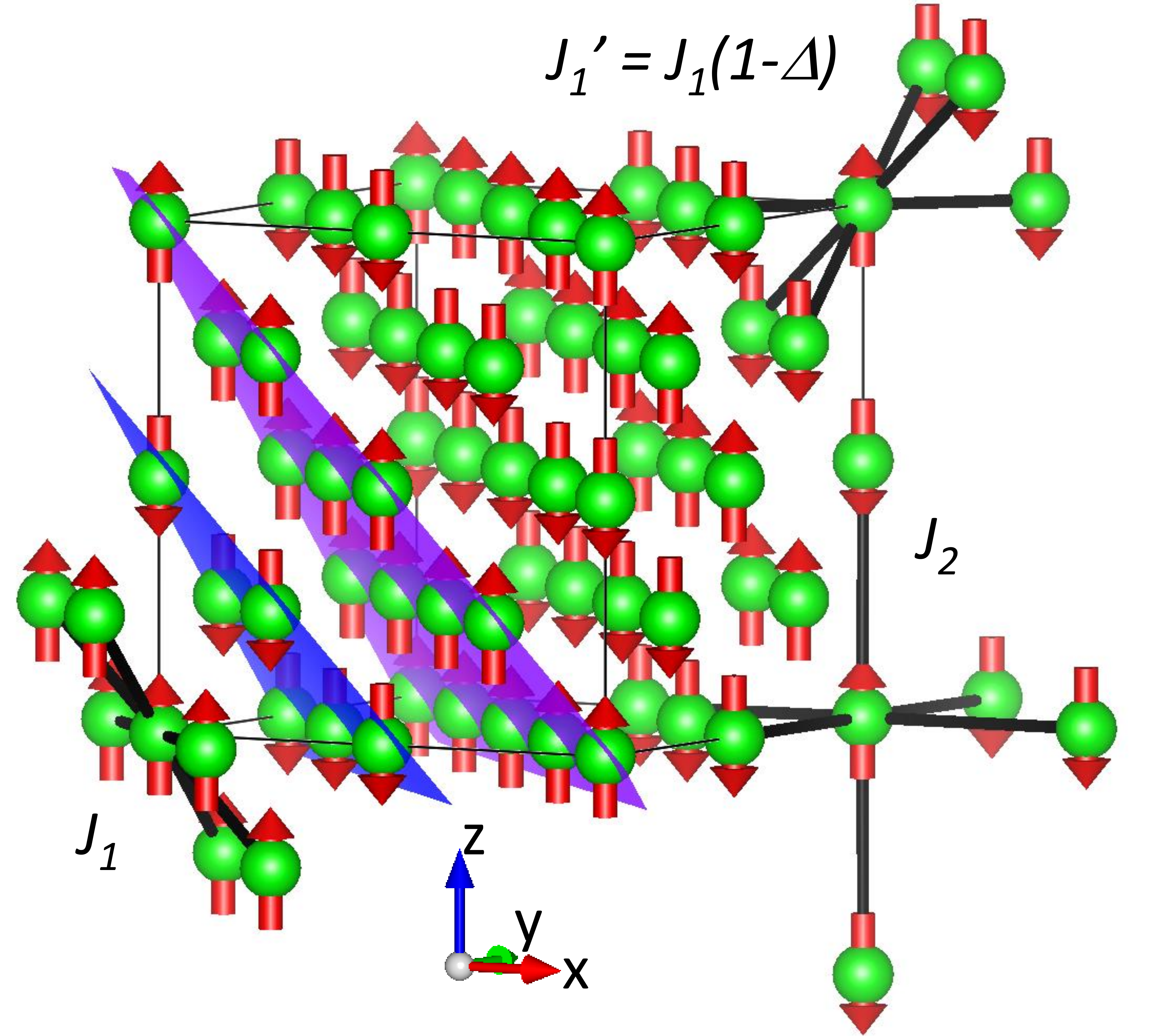}\caption{\label{fig:fig1} Spins on a face centered cubic lattice arranged in the antiferromagnetic order of the second kind with ordering wave vector $q=(1/2,1/2,1/2)$ in which the spins within/between the (111) planes and aligned ferromagnetically/antiferromagnetically. Under the rhombohedral distortion, the nearest neighbor exchange coupling is split into $J_{1}$/$J_{1}'$ between spins within/between the (111) planes, whereas the second nearest neighbor exchange coupling $J_2$ is not split.}
\end{figure}

The magnetic properties of MgO-HEO had remained largely unexplored, until recently. Exchange bias measurements of permalloy/MgO-HEO heterostructures revealed that MgO-HEO is an antiferromagnet~\cite{meisenheimer_2017_srep}. In two subsequent studies, neutron diffraction and magnetic susceptibility measurements have shown that bulk MgO-HEO displays antiferromagnetic (AFM) order with N\'{e}el temperatures reported to be 113 K~\cite{zhang_2019_chemmat} and 120 K~\cite{jimenez_2019_apl}, with magnetic moments arranged in the same AFM ground state as NiO and CoO~\cite{shull_1951,roth_1958}. In this so called AFM order of the second kind, spins are parallel within the (111) planes and antiparallel between the (111) planes (c.f. Fig. ~\ref{fig:fig1}). Inelastic neutron scattering measurements have shown that spin-wave excitations are gapped, with a spin-gap on the order of $\sim 7$ meV~\cite{frandsen_2020_prm}. The gap is gradually buried in quasielastic fluctuations upon heating towards $T_N$. Interestingly, the spin-wave excitations persist above $T_N$, up to room temperature~\cite{zhang_2019_chemmat} . Unlike in binary oxides such as  NiO and CoO, no strong lambda-anomaly has been observed in the specific heat~\cite{zhang_2019_chemmat, jimenez_2019_apl}. Recent muon spin relaxation measurements~\cite{frandsen_2020_prm} indicate that MgO-HEO undergoes a broad continuous AFM phase transition starting at 140 K and being fully ordered at 100 K, explaining why the specific heat anomaly is very weak and broad. The discovery of magnetic long-range order in MgO-HEO leads to a number of questions. How can the experimentally observed AFM long range order be stabilized in the presence of the randomly distributed moments of Co, Ni and Cu and the 40\% of spin vacancies created by the non magnetic Mg and Zn ions? What is the nature of spin-wave excitations in such a extremely disordered magnetic environment? How do the magnetic properties of MgO-HEO compare with those of Ni$_{1-x}$Mg$_x$O~\cite{menshikov_1991}, Ni$_{1-x}$Zn$_x$O~\cite{rodic_2000} and Co$_{1-x}$Mg$_x$O~\cite{seehra_1988}, which also display antiferromagnetic order in the presence of similarly large spin-vacancy concentrations?   

In this paper we theoretically investigate the magnetic ground state and spin-wave excitations in the high entropy oxide MgO-HEO. We model the Co$^{2+}$, Ni$^{2+}$ and Cu$^{2+}$ cations with spins of size $3/2$, $1$ and $1/2$ respectively and the Mg$^{2+}$ and Zn$^{2+}$ cations as being spin-vacancies. We randomly distribute the cations on a rhombohedrally distorted face-centered cubic lattice and analyze the magnetic properties via linear spin-wave theory in combination with the supercell approximation. We find that large next-nearest neighbor exchange couplings and large rhombohedral distortions are required in order to stabilize the experimentally observed antiferromagnetic ground state with propagation vector $q=(1/2,1/2,1/2)$. The spin-wave excitation spectrum of our model for MgO-HEO consists of a coherent component at low energies and an incoherent component at high energies. We also model Co$_{0.33}$Ni$_{0.33}$Cu$_{0.33}$O containing moment-size disorder only and Ni$_{0.6}$Mg$_{0.4}$O containing spin-vacancy disorder only. We find qualitative differences in the spin-wave spectra of these systems compared with that of MgO-HEO. Finally we also investigate the influence of disorder on the spin-wave gap and find that it is proportional to the average moment size per lattice site. 

\section{Methods}

In order to qualitatively investigate the ground state and spin-wave excitations in MgO-HEO we use a simplified model. X-ray absorption spectroscopy~\cite{zhang_2019_chemmat} and extended x-ray absorption fine structure~\cite{rost_2017_jacs} are indicative of the cations in MgO-HEO being divalent. In addition, electron paramagnetic resonance~\cite{jimenez_2019_apl} and Density Functional Theory (DFT)~\cite{rak_2016_jap} have shown that Co$^{2+}$ is in the high spin state. Based on these experimental and theoretical results, we treat the Zn$^{2+}$ and Mg$^{2+}$ cations as spin-vacancies and the Co$^{2+}$, Ni$^{2+}$ and Cu$^{2+}$ cations as spins with moments sizes 3/2, 1 and 1/2 respectively, ignoring orbital contributions here. To describe the interactions between the magnetic moments, we take into account the nearest and next-nearest neighbor exchange couplings. In the rock-salt structure, the nearest neighbor cations are connected via a cation-oxygen-cation group that has a bond angle of 90 degrees. For the second neighboring cations, the mediating oxygen ions are located in between the two cations, such that the cation-oxygen-cation bond angle is 180 degrees. Therefore, according to the Goodenough-Kanamori rules for superexchange~\cite{goodenough_1958, kanamori_1959}, the nearest neighboring spin interactions are ferromagnetic and the next-nearest neighboring interactions are stronger and antiferromagnetic. This is also what has been concluded from DFT calculations for CoO~\cite{deng_apl_2010}, NiO~\cite{kodderitzsch_2002_prb} and MgO-HEO~\cite{rak_2020_jap} and from some inelastic neutron scattering studies for NiO~\cite{hutchings_1972_prb} and CoO~\cite{tomiyasu_2006_jpsj}. Finally, we incorporate a rhombohedral distortion into our model. On an undistorted FCC lattice the antiferromagnetic order of the second kind will be a mixture of four degenerate ordering wave vectors: $q=(1/2,1/2,1/2)$, $q=(-1/2,1/2,1/2)$, $q=(1/2,-1/2,1/2)$ and $q=(1/2,1/2,-1/2)$. A rhombohedral distortion with the trigonal axis along the [111] direction will naturally stabilize the $q=(1/2,1/2,1/2)$ configuration in favor of the other three ordering wave vectors~\cite{yamamoto_1972_jpsj}. Putting it all together, we consider to following model:
\begin{eqnarray}\label{eqn1}
H&=&-J_{1} \sum_{\langle\langle r,r'\rangle\rangle_{1p}} {\bf\hat{ S}}_{r} \cdot {\bf\hat{ S}}_{r'}-J_{1}' \sum_{\langle\langle r,r'\rangle\rangle_{1a}} {\bf\hat{ S}}_{r} \cdot {\bf\hat{ S}}_{r'}
\nonumber \\ &&
-J_{2} \sum_{\langle\langle r,r'\rangle\rangle_{2}} {\bf\hat{ S}}_{r} \cdot {\bf\hat{ S}}_{r'}-K \sum_{r} S_{rz}^2
\end{eqnarray}
where $r$ labels the sites on the disordered cation FCC lattice, $\langle\langle r,r'\rangle\rangle_{1p}$ the first nearest neighbors within the ferromagnetic (111) planes, $\langle\langle r,r'\rangle\rangle_{1a}$ the first nearest neighbors between the ferromagnetic (111) planes and $\langle\langle r,r'\rangle\rangle_{2}$ the second nearest neighbors. Overall our simplified model depends on four parameters, $J_{1}'=J_1(1-\Delta)$, $J_2$ and $K$, where $\Delta$ quantifies the strength of the orthorhombic distortion and $K$ the magnetic anisotropy. This model captures the disorder due to the random distributions of the Co$^{2+}$, Ni$^{2+}$ and Cu$^{2+}$ moment sizes and the Mg$^{2+}$ and Zn$^{2+}$ spin-vacancies. For simplicity, disorder in the magnetic exchange couplings ~\cite{rak_2020_jap} has been ignored. 

To study the spin wave excitation spectra of our model for MgO-HEO, we use a combination of linear spin-wave theory ~\cite{fishman_2018} and the supercell approximation~\cite{berlijn_2011_thesis}. The linear-spin wave formalism consists of two steps. First the classical ground state configuration is determined. To this end we use the conjugate gradient method as implemented in the YaSpinWave computer program~\cite{yaspinwave}. In the second step, the classical ground state is used to perform a Holstein-Primakoff transformation so that our interacting quantum-spin model (\ref{eqn1}) can be approximated as a quadratic boson Hamiltonian that can be solved with exact diagonalization. To treat the strong magnetic disorder in our model for MgO-HEO, we use the supercell approximation in which the disordered system is approximated by a large number of supercells, within which the cations are disordered but beyond which they are artificially periodic. For other studies using linear spin-wave theory in combination with the supercell approximation, we refer to Ref.~\cite{buczek_2016_prb,bai_2019_prl} and for alternative methods to simulate magnetic excitations in disordered systems we refer to Ref. ~\cite{buyers_1973,samarakoon_2017_prb}.  

\section{Ground state}\label{sec2}

Before studying the spin-wave excitations in MgO-HEO, we first focus on the stability of the $q=(1/2,1/2,1/2)$ AFM ground state depicted in Fig. ~\ref{fig:fig1}. In the ordered FCC spin lattice this AFM ground state is stabilized in the parameter regime $|J_2/J_1|>1-\Delta$~\cite{yamamoto_1972_jpsj}. Here $J_1>0$ is the FM nearest and $J_2<0$ the AFM next nearest neighbor exchange couplings respectively, and $\Delta$ parametrizes the strength of the rhombohedral distortion, see Fig. ~\ref{fig:fig1}. From this inequality we can see that the $q=(1/2,1/2,1/2)$ AFM ground state becomes more stable upon increasing the magnitude of the second nearest exchange coupling and the rhombohedral distortion. Figure ~\ref{fig:fig1} shows that intuitively this makes sense.  By inducing a rhombohedral distortion, the magnitude of the FM exchanges within the (111) planes ($J_1$) will increase relatively to those between the (111) planes ($J_{1}'=J_1(1-\Delta)$). The AFM $J_2$ exchange only couples moments between the (111) planes and therefore helps to further stabilize the $q=(1/2,1/2,1/2)$ AFM ground state. 

\begin{figure}[!htb]
\includegraphics[width=1.0\columnwidth]{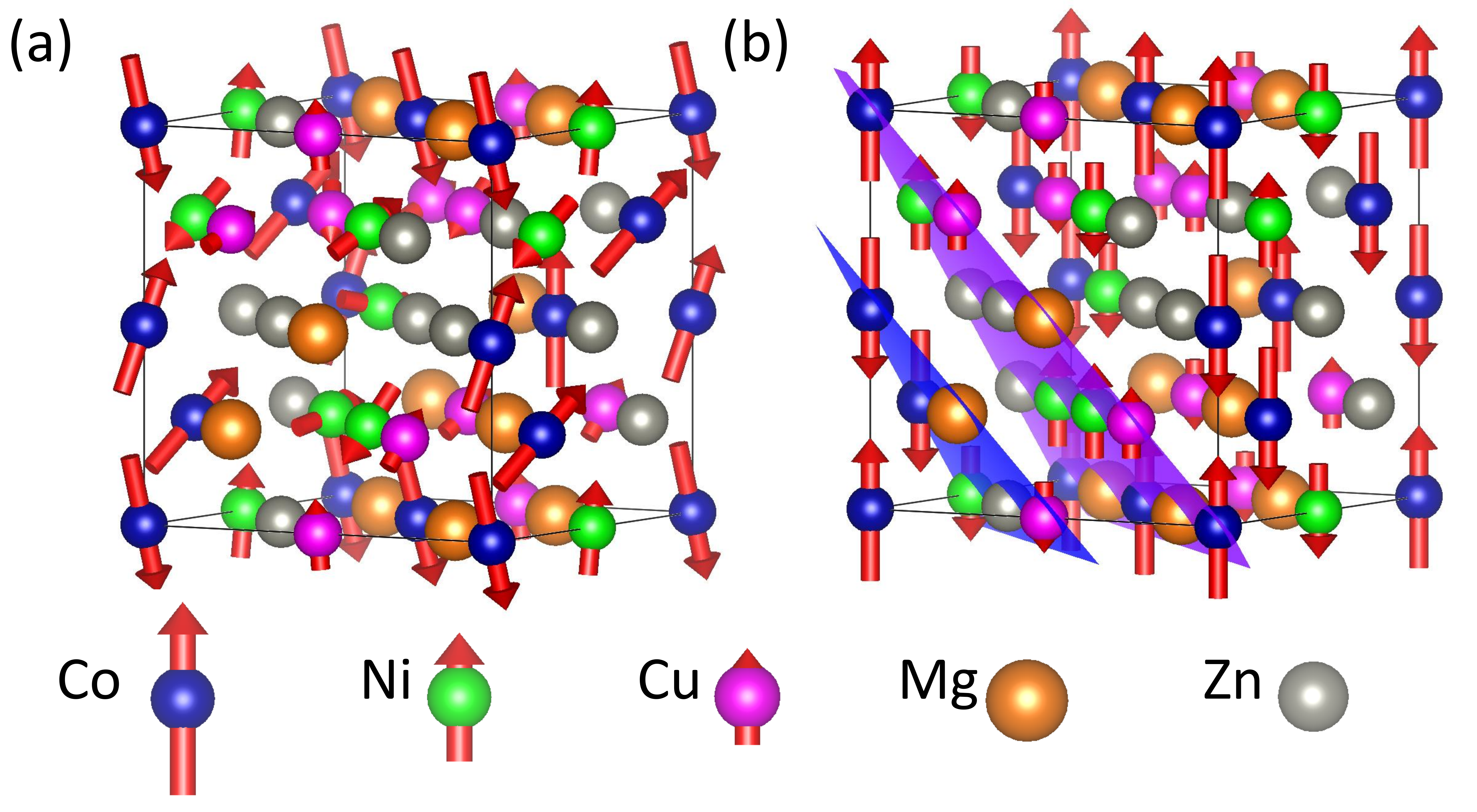}\caption{\label{fig:fig2} Magnetic ground states for a Co$_6$Ni$_6$Cu$_6$Mg$_6$Zn$_8$ supercell in the case of isotropic interactions ($K=0$) with (a) $J_2=-J_1$, $\Delta=0$ and (b) $J_2=-15 J_1$, $\Delta=0.5$.}
\end{figure}

In the presence of disorder the stability arguments for the $q=(1/2,1/2,1/2)$ AFM ground state qualitatively work the same. In Fig. ~\ref{fig:fig2} we consider a $2\times2\times2$ orthogonal supercell relative to the conventional FCC unit cell. This supercell contains 6 $S=3/2$, 6 $S=1$ and 6 $S=1/2$ moments corresponding to the randomly positioned Co, Ni and Cu cations respectively. That leaves 14 randomly distributed spin vacancies corresponding to the Mg and Zn cation sites. Figure ~\ref{fig:fig2}(a) illustrates that for $J_2=J_1$ and $\Delta=0$, the ground state is in some non-collinear configuration. Upon increasing the second nearest exchange to $J_2=-15 J_1$ and the rhombohedral distortion to $\Delta=0.5$, the $q=(1/2,1/2,1/2)$ AFM configuration is stabilized despite the disorder (c.f. Fig. ~\ref{fig:fig2}b). We note that the position of the moments within the supercells in this work have been randomized under the constraint that in the $q=(1/2,1/2,1/2)$ AFM configuration the total moment is zero. To accomplish this we divide the FCC lattice into two sets of parallel (111) planes. The planes for which in the $q=(1/2,1/2,1/2)$ AFM state the moments are up (e.g. the purple plane in Fig. ~\ref{fig:fig2}b) contain an equal amount of Co, Ni and Cu cations as the planes for which in the $q=(1/2,1/2,1/2)$ AFM state the moments are down (e.g. the blue plane in Fig. ~\ref{fig:fig2}b). This implies that the number of Co, Ni and Cu cations per supercell needs to be even. In order to satisfy the equiatomic limit as closely as possible under this constraint, we choose the $2\times2\times2$ FCC supercells to contain 6 magnetic cations of each type and 14 non magnetic cations. 

\begin{table} 
\begin{tabular}{ccccc} \toprule 
             & $J_2=   -J_1$ & $J_2=   -2J_1$ & $J_2=   -5J_1$ & $J_2=  -15J_1$ \\\colrule 
$\Delta=0.0$ &          0\% &          0\% &          0\% &          0\% \\\colrule 
$\Delta=0.1$ &          0\% &          0\% &          2\% &          8\% \\\colrule 
$\Delta=0.2$ &          0\% &          4\% &         17\% &         40\% \\\colrule 
$\Delta=0.5$ &         32\% &         51\% &         67\% &         79\% \\\botrule 
\end{tabular}\caption{\label{tab:tab1} Probability of an orthogonal Co$_6$Ni$_6$Cu$_6$Mg$_6$Zn$_8$ supercell being in the $q=(1/2,1/2,1/2)$ AFM ground state in the case of isotropic interactions: $K=0$, obtained from simulating 100 Co$_6$Ni$_6$Cu$_6$Mg$_6$Zn$_8$ supercells.} 
\end{table}

We now quantitatively investigate the stability of the $q=(1/2,1/2,1/2)$ AFM ground state, by varying the second neighbor exchange coupling and the rhombohedral distortion parameter over a range of values. For each of the $4\times 4$ parameter combinations listed in Table ~\ref{tab:tab1}, we simulate 100 orthogonal disordered Co$_6$Ni$_6$Cu$_6$Mg$_6$Zn$_8$ supercells, again under the constraint that in the q=(1/2,1/2,1/2) AFM configuration the total moment is zero. For each supercell we minimize the classical energy via the conjugate gradient method, starting from five different random spin configurations. The configuration with the lowest energy is assigned to be the ground state. The ground state configuration is typically found in three out of five energy minimizations. In total $4\times 4\times 100\times 5 = 8000$ energy minimizations have been performed to produce Table \ref{tab:tab1}. To determine whether the ground state energy corresponds to the $q=(1/2,1/2,1/2)$ AFM configuration, we considered the average of the absolute difference for each of the spin vectors in these configurations. Appendix ~\ref{appA} contains more details about how we differentiate spin-configurations. 

As expected, the percentage of cases for which the $q=(1/2,1/2,1/2)$ AFM ground state is realized increases as a function of the next nearest exchange coupling $J_2$ and the rhombohedral distortion $\Delta$, see Table ~\ref{tab:tab1}. Yet the strength of the parameters required to stabilize the $q=(1/2,1/2,1/2)$ AFM ground state is considerably higher than for the ordered case, due to the disorder. For example, for the ordered case with $\Delta=0.2$, $J_2=-0.8 J_1$ is sufficient. Our model for the disordered MgO-HEO with $J_2=-15 J_1$ realizes the $q=(1/2,1/2,1/2)$ AFM ground state in only 40\% of the configurations. We have also repeated the same energy minimization in the presence of magnetic anisotropy $K=0.05 J_1$, see Table ~\ref{tab:tab3}. The number of configurations, for which the $q=(1/2,1/2,1/2)$ AFM ground state is realized, remains essentially the same as for the isotropic case, differing at most by a few percentage points compared to the isotropic case shown in Table ~\ref{tab:tab3}.  

\section{Spin-wave excitations}

\begin{figure}[!htb]
\includegraphics[width=1.0\columnwidth]{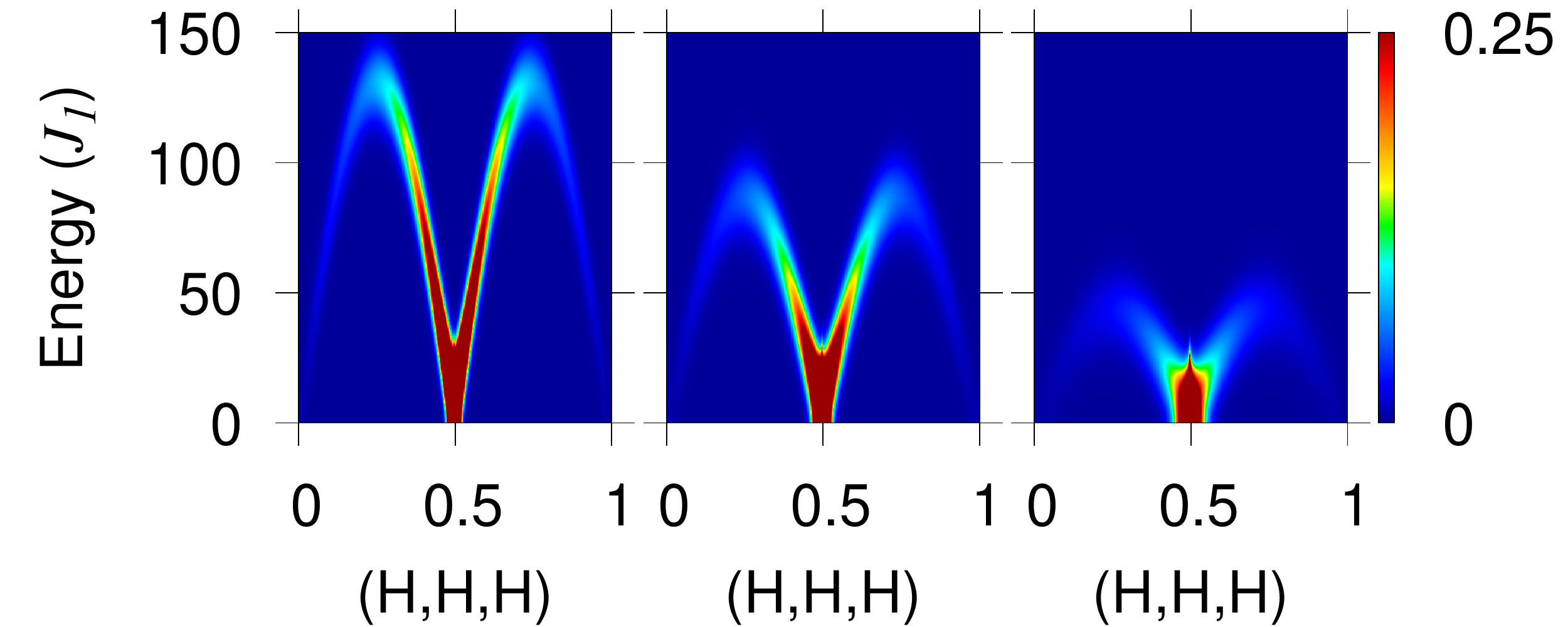}\caption{\label{fig:fig3} Spin-wave spectra of ordered CoO (left), NiO (middle) and CuO (right) FCC lattice spin models with $J_2=-15J_1$, $\Delta=0.5$ and $K=0$.}
\end{figure}

Having understood the stability regime of the classical $q=(1/2,1/2,1/2)$ AFM ground state for our MgO-HEO model, we move on to study the corresponding spin-wave excitation spectrum. To that end we will focus on the parameters $J_2=-15 J_1$ and $\Delta=0.5$ for which the $q=(1/2,1/2,1/2)$ AFM ground state was found to be most stable within our set of parameters. Before analyzing the influence of disorder in our MgO-HEO model, it is instructive to consider the spin-wave spectra of the ordered CoO, NiO and CuO FCC spin models shown in Fig. ~\ref{fig:fig3}. We notice two things. First, the energy range of the spin-wave spectra is proportional to the moment magnitude, S=3/2, 1 and 1/2, of the Co, Ni and Cu cations, respectively. Second, we convoluted the spin-wave spectra with a Gaussian of width $0.5 J_1$ for the purpose of visualization. Other than that, the spin-wave bands in Fig. ~\ref{fig:fig3} are sharp in momentum and energy space.

\begin{figure}[!htb]
\includegraphics[width=1.0\columnwidth]{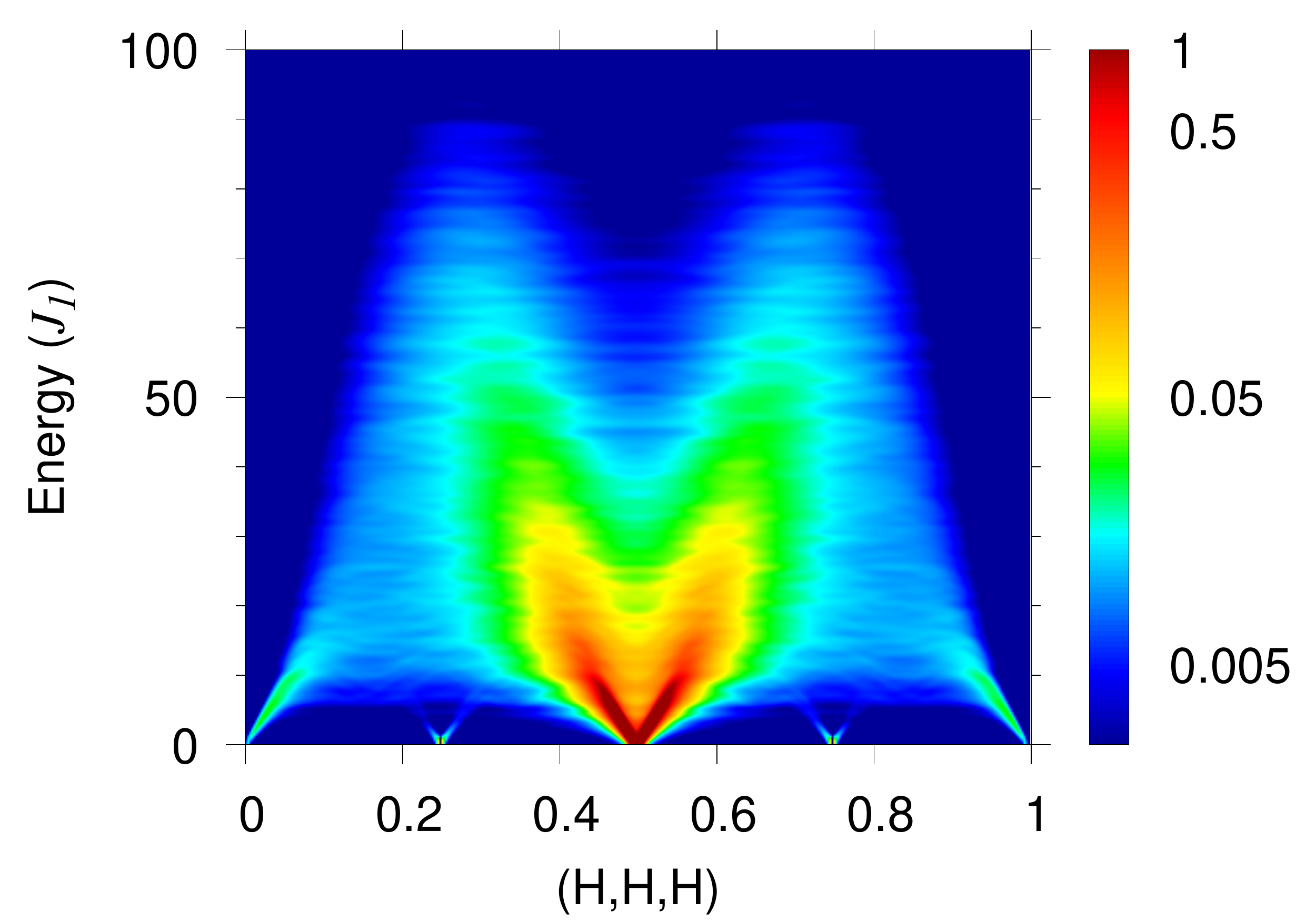}\caption{\label{fig:fig4} Spin-wave spectrum of Mg$_{0.2}$Co$_{0.2}$Ni$_{0.2}$Cu$_{0.2}$Zn$_{0.2}$O FCC lattice spin model with moment size and spin vacancy disorder, obtained from averaging over 100 non-orthogonal supercells with 250 FCC lattice sites on average and $J_2=-15J_1$, $\Delta=0.5$ and $K=0$.}
\end{figure}

\begin{figure}[!htb]
\includegraphics[width=1.0\columnwidth]{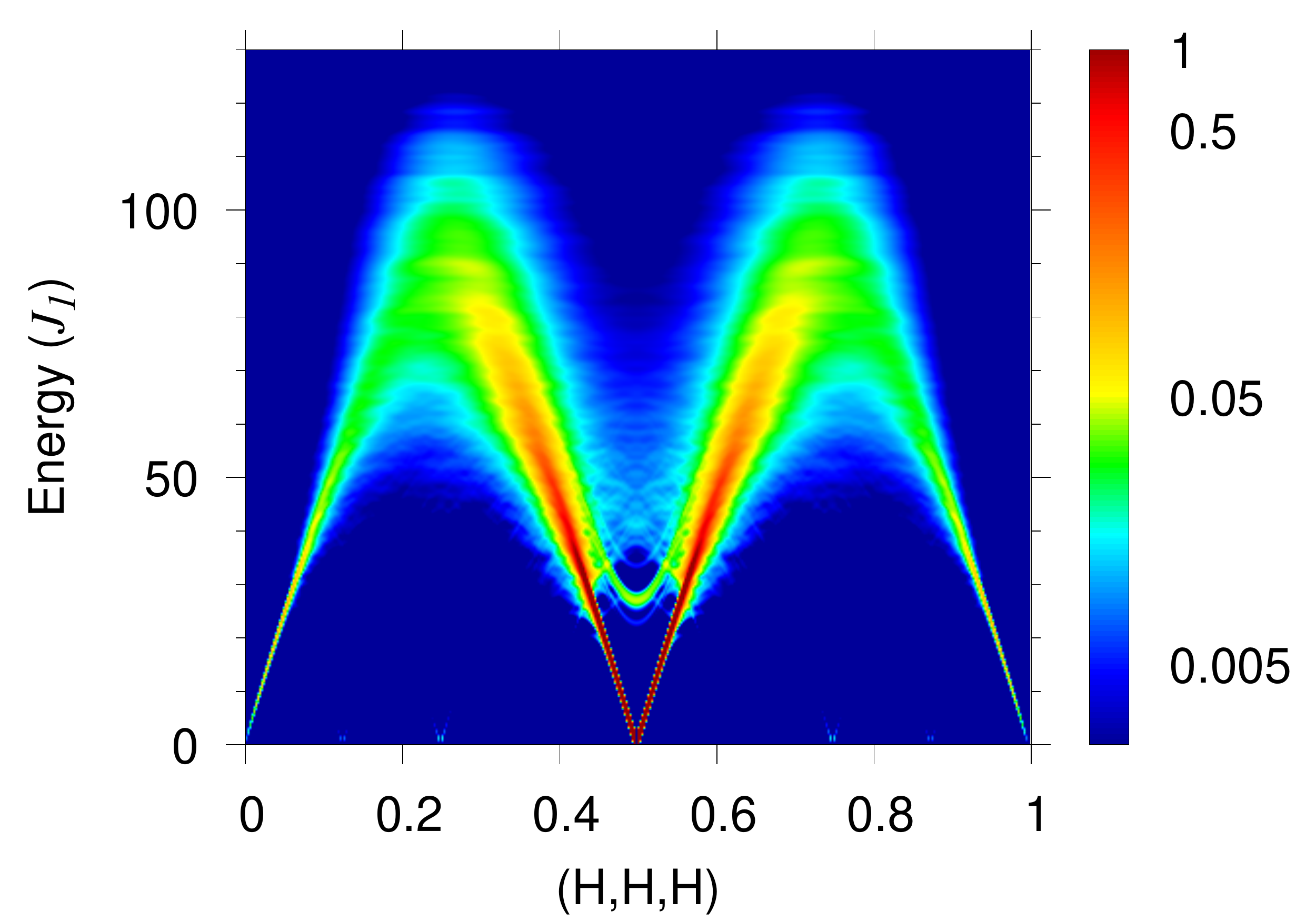}\caption{\label{fig:fig5} Spin-wave spectrum of Co$_{0.33}$Ni$_{0.33}$Cu$_{0.33}$O FCC lattice spin model with moment size disorder and without spin vacancy disorder, obtained from averaging over 100 non-orthogonal supercells with 250 FCC lattice sites on average and $J_2=-15J_1$, $\Delta=0.5$ and $K=0$.}
\end{figure}

\begin{figure}[!htb]
\includegraphics[width=1.0\columnwidth]{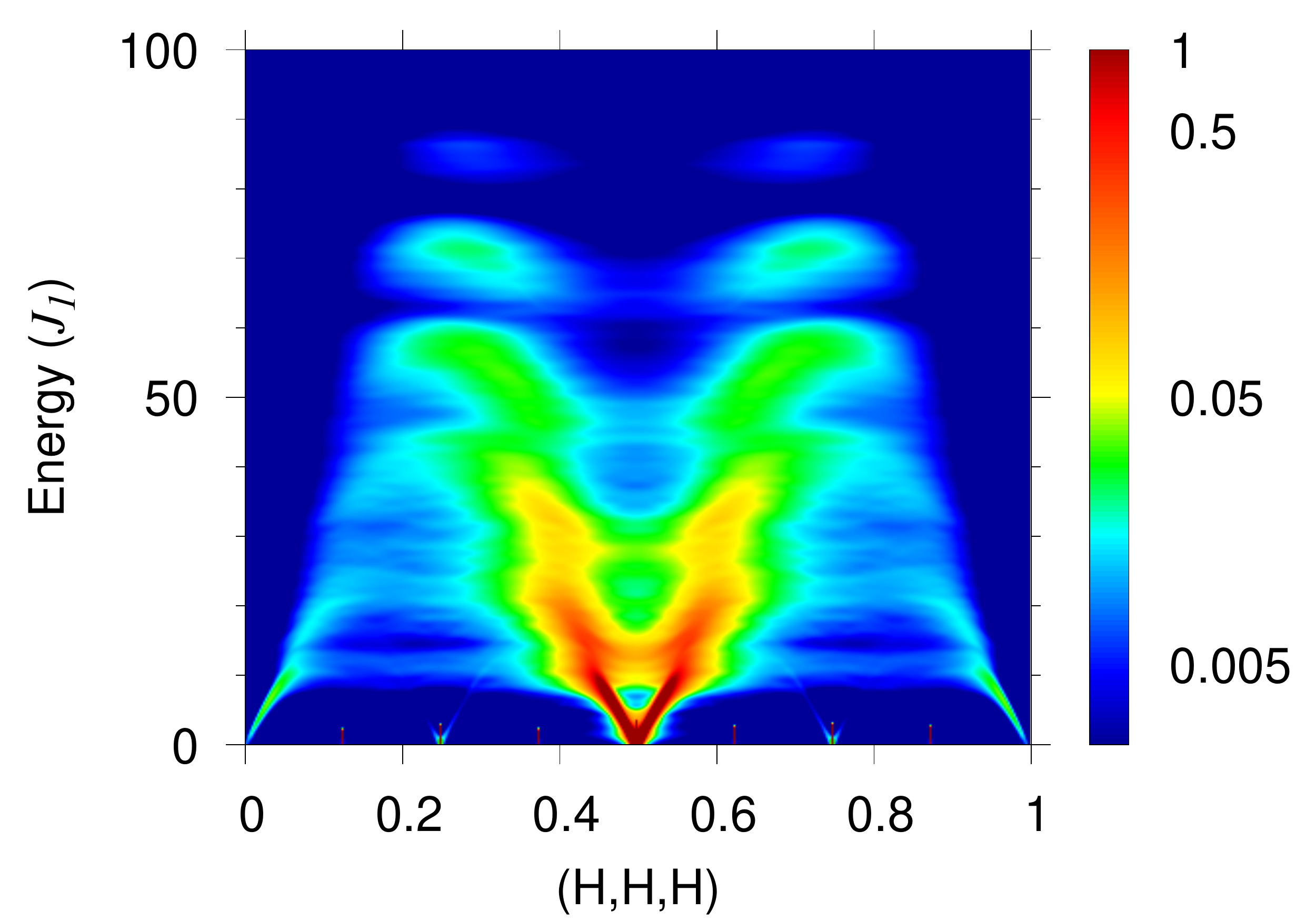}\caption{\label{fig:fig6} Spin-wave spectrum of Ni$_{0.6}$Mg$_{0.4}$O FCC lattice spin model without moment size disorder and with spin vacancy disorder obtained from averaging over 100 non-orthogonal supercells with 250 FCC lattice sites on average and $J_2=-15J_1$, $\Delta=0.5$ and $K=0$.}
\end{figure}

In Fig. ~\ref{fig:fig4} we present the spin-wave spectra for our MgO-HEO model. 
These spectra are obtained from averaging over 100 configurations each containing 250 FCC lattice sites on average over which the Co, Ni and Cu moments and the Mg and Zn spin-vacancies are randomly distributed. We again use the constraint described in the previous section that in the $q=(1/2,1/2,1/2)$ AFM state the total moment vanishes. To better average the disorder configurations, we not only randomly distribute the cations within the supercells, but also randomly change the shapes of the supercells themselves, as App. ~\ref{appB} explains. Comparing the spin-wave spectrum of the strongly disordered MgO-HEO model in Fig.~\ref{fig:fig4} to those of our models for the ordered CoO, NiO and CuO systems in Fig.~\ref{fig:fig3} reveals notable differences. First, the spin-wave spectrum of the MgO-HEO model displays large broadenings in energy and momentum space corresponding to the short lifetimes and short mean-free-paths of the spin-waves due to the presence of strong disorder. Another striking difference is that for the MgO-HEO model the spin-wave spectrum can be characterized by two regions. In the low energy regime, roughly within $[0,10] J_1$, the spin-wave spectrum exhibits a relatively sharp dispersion. At higher energies, roughly within $[10,100] J_1$, the spin-wave spectrum broadens to such an extent, that there is barely any dispersion left. This featureless incoherent part of the spin-wave spectrum should experimentally be difficult to distinguish from the background. Therefore, most likely the observed spin-wave dispersions in inelastic neutron scattering correspond to the coherent part of the spin-wave spectrum in our theoretical calculations. To further illustrate the coherent and incoherent portions of the spin-wave spectrum for the MgO-HEO model, Fig. ~\ref{fig:fig9} in App. ~\ref{appD} presents energy distribution curves for eight fixed momenta near the $q=(1/2,1/2,1/2)$ AFM ordering wave vector. Note that for the spin-wave spectra of our disordered models, see Fig.~\ref{fig:fig4}-\ref{fig:fig6} and \ref{fig:fig9}, a logarithmic scale is used for the spin-wave intensity.      

Next we dissect the consequences of the two different types of disorder in our MgO-HEO model, the moment size disorder induced by the Co, Ni and Cu cations and the spin-vacancy disorder induced by the Mg and Zn cations. In Fig.~\ref{fig:fig5} we focus on the moment size disorder by presenting the spin-wave spectrum of the Co$_{0.33}$Ni$_{0.33}$Cu$_{0.33}$O model in which there is only moment size disorder, but no spin-vacancy disorder. We note two differences between Fig.~\ref{fig:fig5} and Fig.~\ref{fig:fig4}. First, the low energy part of the spectrum of the Co$_{0.33}$Ni$_{0.33}$Cu$_{0.33}$O model is much sharper than for our MgO-HEO model. This illustrates that the scattering of low energy spin-waves against moment size disorder is strongly suppressed as a function of energy. The situation is analogous to case of low energy acoustic phonons scattering against mass disorder, for which in 3D the mass perturbation is proportional to the quadratic power of the energy; see Ref. ~\cite{thebaud_2020} and references therein. A second difference is that the magnon propagation velocity, that is, the slope of the low-energy spin-wave band close to the $q=(1/2,1/2,1/2)$ AFM ordering wave vector, is significantly higher. Similarly, the maximum energy of the spin-wave intensity is also higher. The reason for the reduction of the magnon propagation velocity and the maximum energy of the spin-wave intensity in the MgO-HEO is the presence of the spin vacancies, which make it energetically less costly for the remaining spins to rotate away from the AFM ground state.

In Fig. ~\ref{fig:fig6} the spin-wave excitation spectrum for the Ni$_{0.6}$Mg$_{0.4}$O model is presented. In this model there is an equal concentration of disordered spin-vacancies as in the MgO-HEO model, but no moment size disorder. The magnon propagation velocities and the maximum energy of the spin-wave intensity in Fig. ~\ref{fig:fig4} and ~\ref{fig:fig6} are quite similar. Moreover, the spin-wave excitation spectrum for the Ni$_{0.6}$Mg$_{0.4}$O model is coherent roughly within $[0,10] J_1$ and incoherent within $[10,100] J_1$, just as for the MgO-HEO model.  One conspicuous difference is that the incoherent part of the spin-wave excitation spectrum of Ni$_{0.6}$Mg$_{0.4}$O shows a number of strong gaps and kinks. Diluted FCC antiferromagnets have been realized experimentally. For example, $q=(1/2,1/2,1/2)$ AFM order in the presence of large spin-vacancy concentrations has been observed in Ni$_{1-x}$Mg$_x$O~\cite{menshikov_1991}, Ni$_{1-x}$Zn$_x$O~\cite{rodic_2000} and Co$_{1-x}$Mg$_x$O~\cite{seehra_1988}. It would be interesting if the kinks and gaps in our Ni$_{0.6}$Mg$_{0.4}$O simulations could be observed experimentally. The nature of the kinks and gaps in the Ni$_{0.6}$Mg$_{0.4}$O is possibly related to the different distributions of Mg spin vacancies around the local Ni spins. We will leave this question for future investigations.  

\begin{table}[!htb] 
    \begin{tabular}{l@{\hspace{1cm}}c}\toprule 
             & spin-gap ($J_1$) \\\colrule 
CoO, NiO, CuO &          6.26,4.17,2.09  \\\colrule 
Co$_{0.2}$Ni$_{0.2}$Cu$_{0.2}$Mg$_{0.2}$Zn$_{0.2}$O &         2.43  \\\colrule 
Co$_{0.33}$Ni$_{0.33}$Cu$_{0.33}$O &         4.29  \\\colrule 
Ni$_{0.6}$Mg$_{0.4}$O &         2.57  \\\botrule 
\end{tabular}\caption{\label{tab:tab2} Spin-gaps induced by easy-axes anisotropy $K=0.05 J_1$ for ordered and disordered FCC lattice spin model with various cation compositions with $J_2=-15J_1$ and $\Delta=0.5$. } 
\end{table}

Finally, we investigate the influence of magnetic disorder on the spin-wave gap. Table ~\ref{tab:tab2} lists the spin-gaps induced by an easy-axes anisotropy $K=0.05 J_1$ for various model systems. For the CoO, NiO and CuO models without disorder, we see that the spin-wave gap is proportional to the size of the cation moments, as expected. To extract the spin-gap for our disordered models, we fit the spin-wave excitation spectrum at the $q=(1/2,1/2,1/2)$ AFM ordering wave-vector with a Gaussian. The spin-wave gap that we obtain for Co$_{0.33}$Ni$_{0.33}$Cu$_{0.33}$O is nearly equal to that for our NiO model. This suggest that the the spin-wave gap is proportional to the average moment size: $\frac{1}{3}(3/2+1+1/2)=1$. This is further confirmed by the fact that the spin-wave gap of the MgO-HEO and the Ni$_{0.6}$Mg$_{0.4}$O model are close in energy. For these two models we also notice that the spin-gap is roughly proportional to $(1-x)$ with $x$ the vacancy concentration. For example, the spin-wave gap of NiO multiplied by the moment concentration is $0.6\cdot4.17=2.51$, i.e. approximately equal to the spin-wave gap of Ni$_{0.6}$Mg$_{0.4}$O of 2.57, and of Mg-HEO of 2.43. Overall, we find that the spin-gap in our model is roughly proportional to the average moment size if we count the spin-vacancies as having zero magnetic moment. 

\section{Discussion}

In our study we have qualitatively modeled the magnetic ground state and spin-wave excitations in MgO-HEO with a rhombohedrally distorted FCC spin model that depends on four global parameters and fully takes into account the moment size and spin-vacancy disorder. One aspect not modeled here and to be considered in future studies is the influence of disorder in the exchange couplings on the spin-wave excitations.  Recently, the exchange couplings in MgO-HEO have been computed from first principles~\cite{rak_2020_jap}. From fitting the total energy of 3 magnetic configurations in 6 ordered cation distributions the ratio of the second to first nearest neighbor exchange couplings was found to vary from $J_2=-11.4J_1$ for Ni-Ni pairs to $J_2=-2.6J_1$ for Cu-Cu pairs, with an average ratio of $J_2=-6.6J_1$. This is reasonably consistent with our conclusion that $J_2=-5J_1$ is needed to stabilize a significant portion of the disordered configurations in the $q=(1/2,1/2,1/2)$ AFM ground state. In this same study ~\cite{rak_2020_jap} an average fitting of the exchange coupling was performed via calculations in a large disordered supercell and led to the conclusion that $J_2=-15.1J_1$, very close to the value used in our study to study spin-wave excitations. Another aspect that would be significant to consider is the influence of local distortions on the exchange couplings. Local Jahn-Teller distortions of the oxygen anions surrounding the Cu$^{2+}$ cations have been observed in MgO-HEO via extended X-ray absorption fine structure spectroscopy~\cite{rost_2017_jacs},  electron paramagnetic resonance and X-ray diffraction studies~\cite{berardan_2017_jac}. From DFT~\cite{rak_2018_ml} it is concluded that these Jahn-Teller distortions are randomly orientated. Such distortions could induce a large spread in the exchange couplings even for first or second nearest neighbor couplings of a fixed inter-species pair. For example, for the high entropy alloy CrFeCoNi, DFT calculations show that the nearest neighbor Fe-Fe interactions can fluctuate roughly between $+10$ and $-10$ meV depending on the local chemical environment~\cite{fukushima_2017_jpsj}. A realistic spin model can be bench-marked against inelastic neutron scattering. Features that can be used to constrain the model are the experimentally observed spin-wave gap of 7 meV ~\cite{frandsen_2020_prm} and the magnon propagation velocity ~\cite{zhang_2019_chemmat}. In addition, our current study has found that the spin-wave excitations in MgO-HEO consist of a coherent part and a featureless incoherent part. Inelastic neutron scattering experiments on MgO-HEO are most likely only able to resolve the coherent portion of the spin-wave spectrum. The maximum of the coherent spectra (in our current simulations roughly equal to $10 J_1$) is an additional quantitative aspect of the spin-wave spectrum that can be benchmarked against inelastic neutron scattering.

Another interesting aspect to consider in future investigations is the magnetic frustration induced by competition between the first and second nearest neighbor exchange couplings. In this study we have worked under the assumption that below the N\'{e}el temperature MgO-HEO orders in the $q=(1/2,1/2,1/2)$ AFM ground state. This is what has been reported based on neutron diffraction studies on MgO-HEO and it is also the ground state that has been experimentally observed for the end members NiO and CoO~\cite{shull_1951,roth_1958}. However, it is possible that no global rhombohedral distortion takes place. Indeed, as of yet, no global distortion has been observed experimentally, although this could also be due to limited resolution. In the absence of a rhombohedral distortion, the AFM ground state of MgO-HEO could be a mixture of the four ordering wave vectors $q=(1/2,1/2,1/2)$, $q=(-1/2,1/2,1/2)$, $q=(1/2,-1/2,1/2)$ and $q=(1/2,1/2,-1/2)$ as is the case for the ordered FCC model with weak nearest FM and strong next nearest AFM couplings ~\cite{yamamoto_1972_jpsj}. Note that such a ground state is not a mixture of single $q$ domains; even within the domains the four $q$ vectors are mixed. The AFM ground state has been resolved from powder diffraction for which it is not possible to distinguish between the single or the mixed $q$ state. Interestingly, in the absence of a rhombohedral distortion, the first nearest neighbor interaction is frustrated. As can be seen in Fig. ~\ref{fig:fig1}, half of the nearest FM exchange couplings, labeled $J_{1}'$, are frustrated in the $q=(1/2,1/2,1/2)$ configuration. A moderate frustration, expressed as the ratio of the Curie-Weiss temperature over the Neel temperature, has in fact been reported for MgO-HEO ~\cite{jimenez_2019_apl}. Frustration of the nearest neighbor interaction could possibly also explain some of the other curious magnetic properties of MgO-HEO, such as for example the broad phase transition ranging from 140 to 100 K~\cite{frandsen_2020_prm}, the persistence of spin-wave excitations up to room temperature~\cite{zhang_2019_chemmat}, and the bifurcation of the field-cooled and zero field cooled magnetic susceptibility~\cite{zhang_2019_chemmat,jimenez_2019_apl}. 

\section{Conclusion}

In this study we have investigated the spin-wave excitations in the high entropy oxide Mg$_{0.2}$Co$_{0.2}$Ni$_{0.2}$Cu$_{0.2}$Zn$_{0.2}$O (MgO-HEO). To model the magnetic interactions in MgO-HEO, we utilize a FCC spin model with FM nearest, and AFM next nearest neighbor interactions in the presence of a rhombohedral distortion and magnetic anisotropy. The Co, Ni and Cu cations are modeled as spins with a moment size of $3/2$, $1$ and $1/2$ respectively, whereas the Mg and Zn cations are taken to be spin-vacancies.  We employ linear spin-wave theory in combination with the supercell approximation to treat the disorder in order to theoretically investigate the ground state and the spin-wave excitations. Stabilizing the experimentally reported $q=(1/2,1/2,1/2)$ AFM ground state in our disordered model for MgO-HEO requires significantly larger second nearest neighbor interactions and rhombohedral distortions compared to the case without disorder. The calculated spin-wave spectrum of MgO-HEO consists of a coherent dispersive part at low energies, whereas at higher energies the spectrum is incoherent due to the disorder induced broadening. To differentiate the influence of moment size disorder and the spin-vacancy disorder we compute the spin-wave spectrum of Co$_{0.33}$Ni$_{0.33}$Cu$_{0.33}$O and Ni$_{0.6}$Mg$_{0.4}$O and find these to be qualitatively different from each other and from the one of MgO-HEO. We also investigated the spin-wave gap for various cation compositions and find that for the considered model it is proportional to the average moment size per lattice site.

\section{Acknowledgments}

TB, DP, RH and RF were supported by the U.S. Department of Energy, Office of Science, Basic Energy Sciences, Materials Sciences and Engineering Division. We used resources of the National Energy Research Scientific Computing Center, a DOE Office of Science User Facility supported by the Office of Science of the U.S. DOE under Contract No. DE-AC02-05CH11231. GA was supported by the scientific Discovery through Advanced Computing (SciDAC) program funded by U.S. Department of Energy, Office of Science, Advanced Scientific Computing Research and Basic Energy Sciences, Division of Materials Sciences and Engineering. Part of this research was conducted at the Center for Nanophase Materials Sciences, which is a DOE Office of Science User Facility.

This manuscript has been authored by UT-Battelle, LLC under Contract No. DE-AC05-00OR22725 with the U.S. Department of Energy. The United States Government retains and the publisher, by accepting the article for publication, acknowledges that the United States Government retains a non-exclusive, paid-up, irrevocable, world-wide license to publish or reproduce the published form of this manuscript, or allow others to do so, for United States Government purposes. The Department of Energy will provide public access to these results of federally sponsored research in accordance with the DOE Public Access Plan (http://energy.gov/downloads/doe-public-access-plan).

\appendix

\section{Differentiating spin configurations}\label{appA}

To quantify how close the ground state of a disordered supercell is to the $q=(1/2,1/2,1/2)$ AFM configuration one could compute the $q=(1/2,1/2,1/2)$ AFM energy relative to the ground state energy. However, for comparing among different parameter sets the energy isn't a good measure, because for a given configuration the energy varies as a function of the parameters. Therefore instead we use the spin-configuration of the ground state to determine its closeness to $q=(1/2,1/2,1/2)$ AFM configuration. To this end we define the difference between two spin configurations as the  absolute difference of the spin-vectors averaged over the spins in the configurations. While doing this, we have to be mindful of the rotational invariance when computing this difference. For example, in the isotropic case the difference between two spin-configurations should not depend on an overall rotation of either of the spin-configurations. To this end we define the difference between spin-configurations $A$ and $B$ as follows. We first perform an overall rotation of configuration $B$ such that its first spin aligns as much as possible with the first spin of configuration $A$ and then compute the average absolute difference of the spin-vectors. To not bias our measure of difference toward the first spin, we also repeat the procedure with aligning each of the other spins as much as possible. Following the above, the difference of spin-configurations $B$ and $A$ is given by:
\begin{eqnarray}
\Delta S^{A,B}=\frac{1}{N^2}\sum_{r,r'} |S^A_r-S^{A,B}_{r,r'}|
\end{eqnarray}
where $N$ is the total number of spins, $S^A_r$ is the spin-vector of spin $r$ in configuration $A$ and 
\begin{eqnarray}
S^{A,B}_{r,r'}=U^{A,B}_{r'} \cdot S^B_r
\end{eqnarray}
is the spin-vector of spin $r$ in configuration $B$ that is rotated to align its spin $r'$ as much as possible with the corresponding spin $r'$ in configuration $A$. For isotropic interactions, the rotation matrix is an element of the 3D rotation group: $U^A_{r'} \in SO(3)$. In that case, the spin $r'$ of configuration A and B can be perfectly aligned. In the presence of an easy axis anisotropy $K>0$, the rotation matrix is an element of the product of the group of 2D rotations around the easy axis and the group of reflections in the plane perpendicular to the easy axis  : $U^A_{r'} \in SO(2)\times Z_2$. For example, suppose we have an easy axis along the z-direction and we have two spins: $S_A=(\cos(\phi_A)\sin(\theta_A), \sin(\phi_A)\sin(\theta_A),\cos(\theta_A))$ and $S_B=(\cos(\phi_B)\sin(\theta_B), \sin(\phi_B)\sin(\theta_B),\cos(\theta_B))$. In that case the rotation matrix $U^{A,B}$ rotates $S_B$ to $(\cos(\phi_A)\sin(\theta_B'), \sin(\phi_A)\sin(\theta_B'),\cos(\theta_B'))$ with $\theta_B'$ such that $\sgn(\cos(\theta_B'))=\sgn(\cos(\theta_A))$ and $|\cos(\theta_B')|=|\cos(\theta_B)|$. 

Having defined a measure to differentiate two spin-configurations, we have to choose a cut-off below which we consider two spin-configurations to be equal. In this work, we consider a spin-configuration $A$ to  have the $q=(1/2,1/2,1/2)$ AFM ground state if  $\Delta S<10^{-4}$. For example, the configuration in Fig.~\ref{fig:fig2}(b) (Fig.~\ref{fig:fig2}(a)) has $\Delta S$ equal to 0.00002(2.12636) and therefore is considered to be (not to be) in the $q=(1/2,1/2,1/2)$ AFM ground state. 

\section{Non-orthogonal supercells}\label{appB}

\begin{figure}[!htb]
\includegraphics[width=0.8\columnwidth]{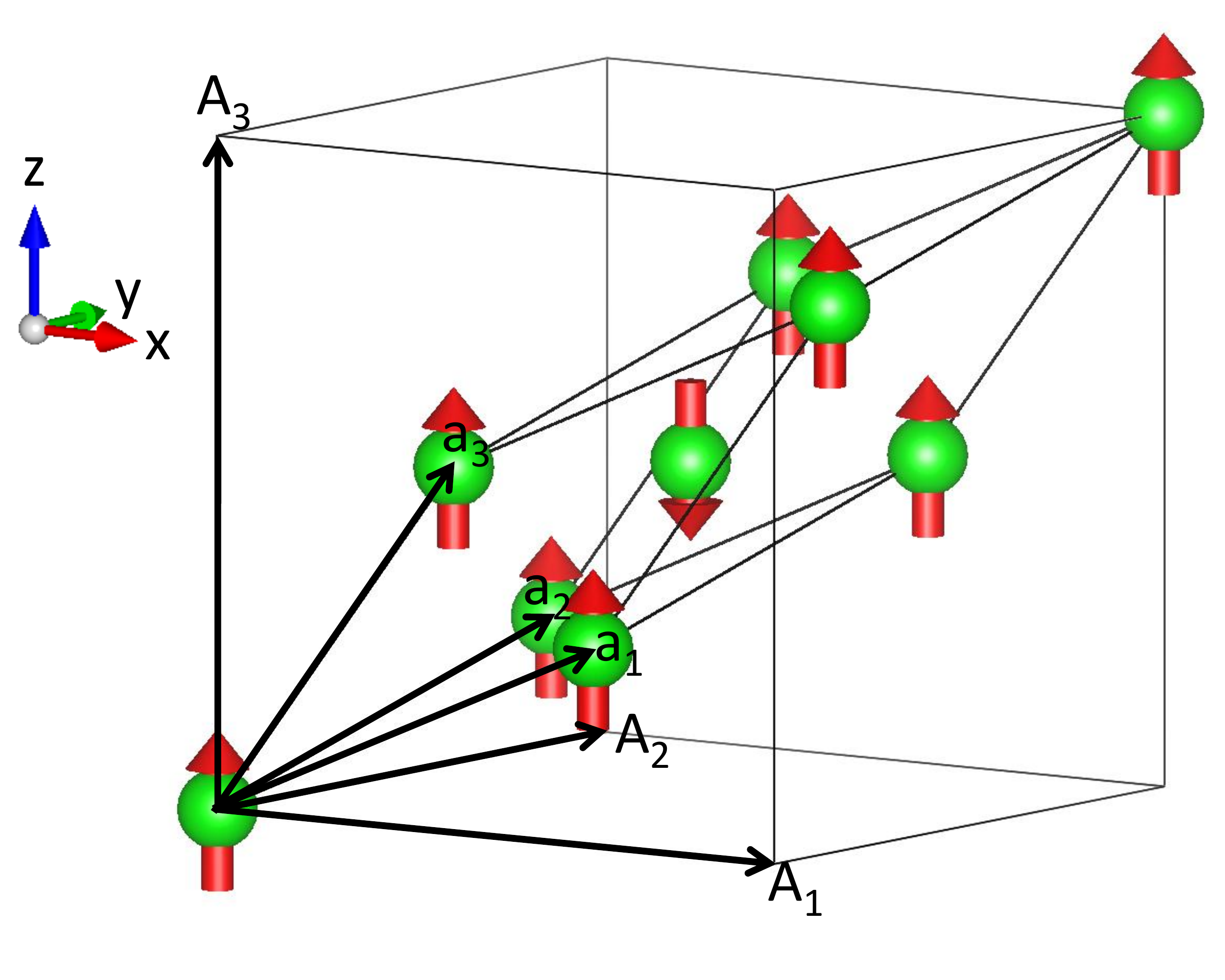}\caption{\label{fig:fig7} Primitive cell of the $q=(1/2,1/2,1/2)$ AFM configuration spanned by $a_1$,$a_2$ and $a_3$ imbedded in the corresponding orthogonal conventional cell spanned by $A_1=3a_1-a_2-a_3$, $A_2=-a_1+3a_2-a_3$ and $A_3=-a_1-a_2+3a_3$.}
\end{figure}

To better average the disorder configurations we don't only randomly distribute the cations within the supercells, but also randomly change the shapes of the supercells themselves~\cite{berlijn_2011_thesis}. To construct the non-orthogonal supercells, we first consider the primitive cell of the antiferromagnetic order of the second kind with ordering wave vector $q=(1/2,1/2,1/2)$ illustrated in Fig.~\ref{fig:fig7}. 

\begin{figure}[!htb]
\includegraphics[width=1.0\columnwidth]{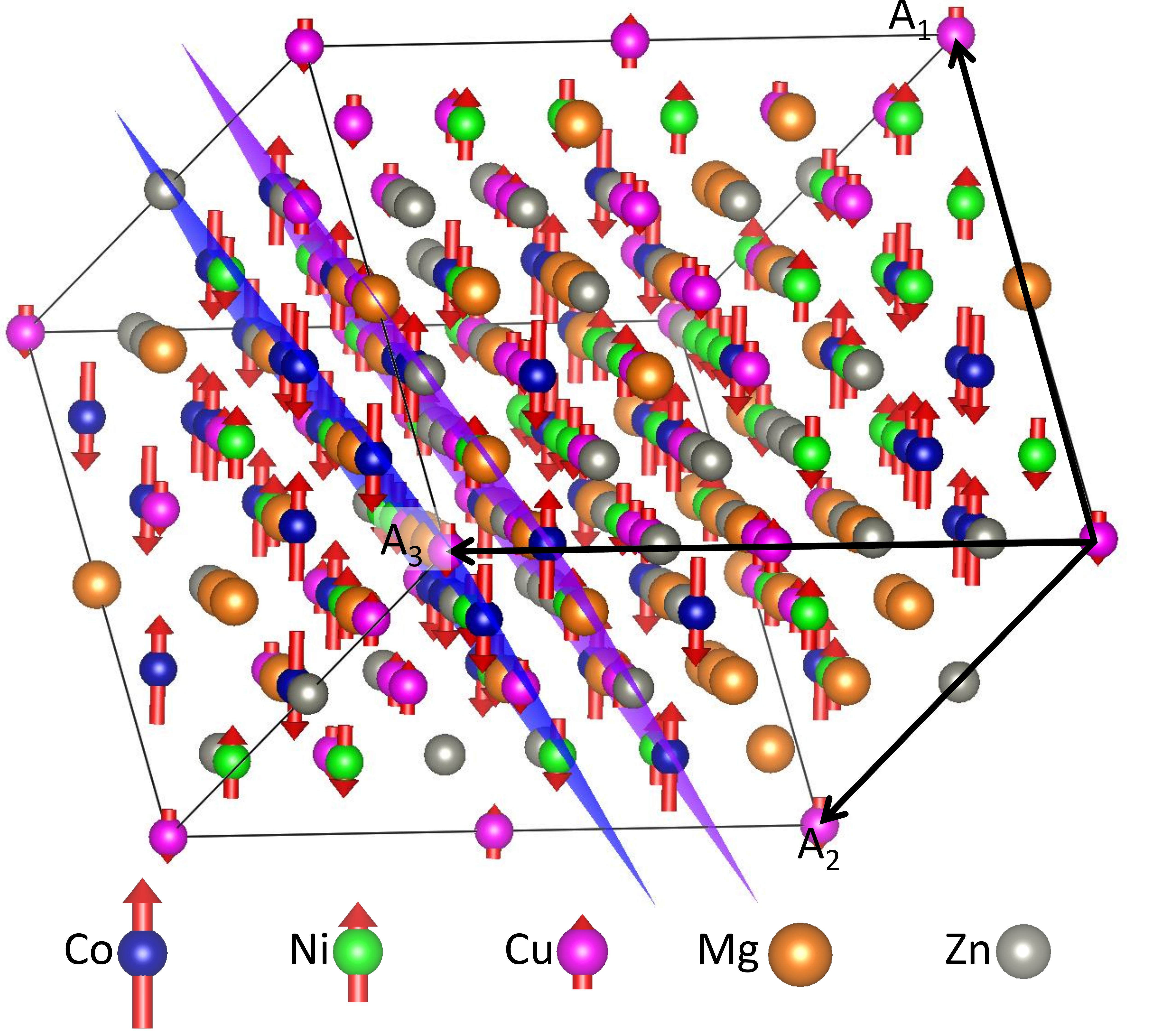}\caption{\label{fig:fig8} One of the non-orthogonal supercells used for computing the disorder-configuration averaged spin-wave spectra.}
\end{figure}

We construct non-orthogonal supercells that are supercells relative to the primitive cell of the $q=(1/2,1/2,1/2)$ AFM configuration. We note that while these resulting non-orthogonal supercells satisfy the boundary conditions of the $q=(1/2,1/2,1/2)$ AFM configuration, they do not satisfy the boundary conditions of the $q=(-1/2,1/2,1/2)$, $q=(1/2,-1/2,1/2)$ and $q=(1/2,1/2,-1/2)$ supercells. As a result, the $q=(1/2,1/2,1/2)$ AFM configuration is significantly more stable in these non-orthogonal supercells compared to the orthogonal ones used to determine the ground state configuration in section ~\ref{sec2}. We select the sizes of these supercells to be within 115 and 135 $q=(1/2,1/2,1/2)$ AFM primitive cells, corresponding to 230 and 270 cation sites. We allow the unit cell angles to vary between 80 and 100 degrees. In Fig. ~\ref{fig:fig8} we illustrate one of the non-orthogonal supercells used for computing the disorder-configuration averaged spin-wave spectra. It is spanned by $A_1=-a_1+5a_2-5a_3$, $A_2=-4a_1+4a_2+2a_3$ and $A_3=5a_1+a_2-3a_3$. It contains 248 cations and its unit cell angles are $\alpha=83.1354$, $\beta=90$ and $\gamma=85.904$. 

\section{Additional ground state results}\label{appC}

\begin{table}[!htb] 
\begin{tabular}{ccccc}\toprule 
             & $J_2=   -J_1$ & $J_2=   -2J_1$ & $J_2=   -5J_1$ & $J_2=  -15J_1$ \\\colrule 
$\Delta=0.0$ &          0\% &          0\% &          0\% &          0\% \\\colrule 
\hline 
$\Delta=0.1$ &          0\% &          0\% &          5\% &          9\% \\\colrule 
$\Delta=0.2$ &          1\% &          6\% &         12\% &         66\% \\\colrule 
$\Delta=0.5$ &         39\% &         57\% &         70\% &         82\% \\\colrule 
\end{tabular}\caption{\label{tab:tab3} Probability of an orthogonal Co$_6$Ni$_6$Cu$_6$Mg$_6$Zn$_8$ supercell being in the $q=(1/2,1/2,1/2)$ AFM ground state in the case of anisotropic interactions: $K=0.05 J_1$, obtained from simulating 100 Co$_6$Ni$_6$Cu$_6$Mg$_6$Zn$_8$ supercells.} 
\end{table}


\section{Additional spin-wave excitation result}\label{appD}

\begin{figure}[!htb]
\includegraphics[width=1.0\columnwidth]{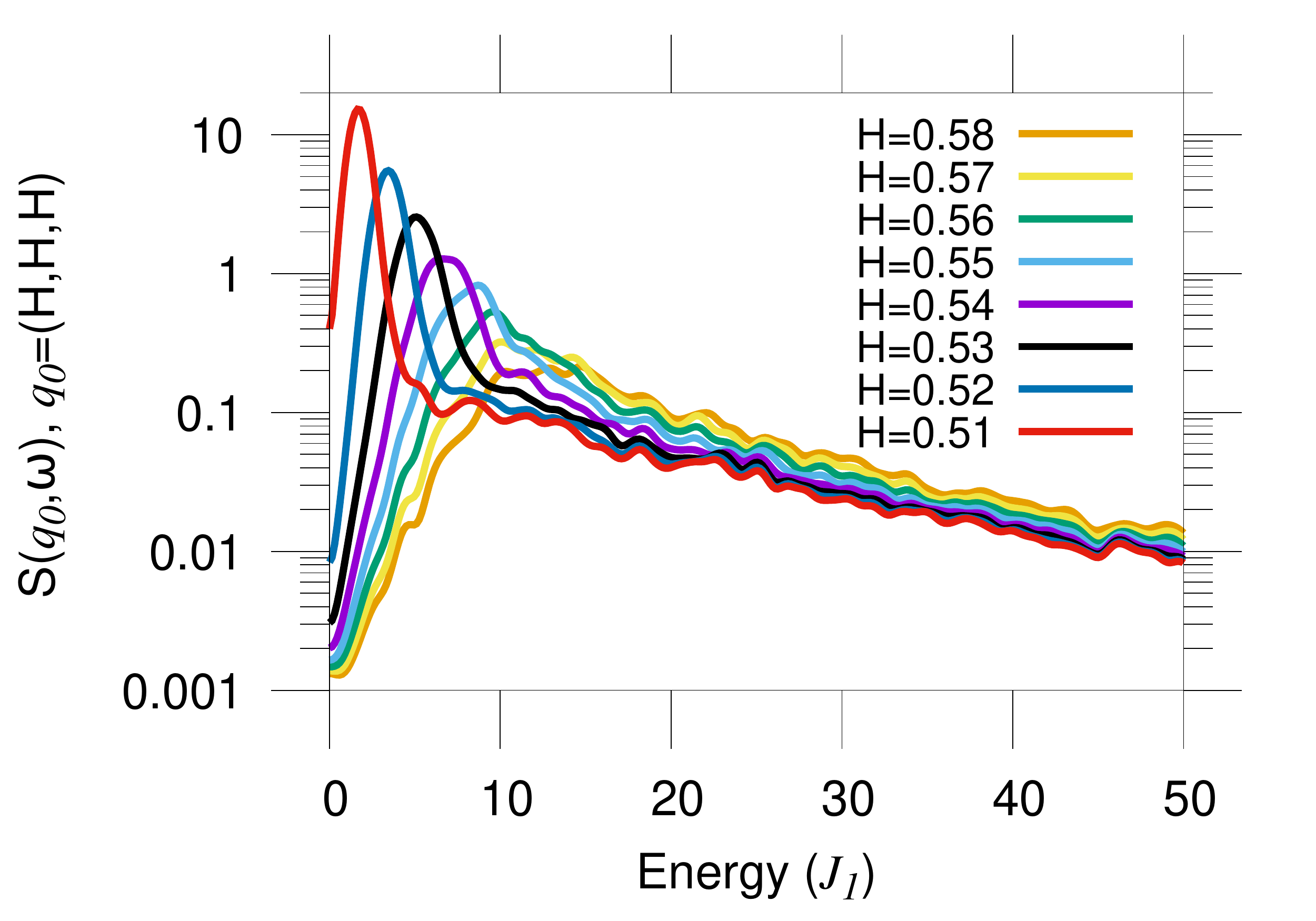}\caption{\label{fig:fig9} Energy distribution curves of the spin-wave spectrum of disordered Mg$_{0.2}$Co$_{0.2}$Ni$_{0.2}$Cu$_{0.2}$Zn$_{0.2}$O FCC lattice spin model obtained from averaging over 100 non-orthogonal supercells with 250 FCC lattice sites on average and $J_2=-15J_1$, $\Delta=0.5$ and $K=0$.}
\end{figure}


\bibliography{reference}

\begin{thebibliography}{46}%
\makeatletter
\providecommand \@ifxundefined [1]{%
 \@ifx{#1\undefined}
}%
\providecommand \@ifnum [1]{%
 \ifnum #1\expandafter \@firstoftwo
 \else \expandafter \@secondoftwo
 \fi
}%
\providecommand \@ifx [1]{%
 \ifx #1\expandafter \@firstoftwo
 \else \expandafter \@secondoftwo
 \fi
}%
\providecommand \natexlab [1]{#1}%
\providecommand \enquote  [1]{``#1''}%
\providecommand \bibnamefont  [1]{#1}%
\providecommand \bibfnamefont [1]{#1}%
\providecommand \citenamefont [1]{#1}%
\providecommand \href@noop [0]{\@secondoftwo}%
\providecommand \href [0]{\begingroup \@sanitize@url \@href}%
\providecommand \@href[1]{\@@startlink{#1}\@@href}%
\providecommand \@@href[1]{\endgroup#1\@@endlink}%
\providecommand \@sanitize@url [0]{\catcode `\\12\catcode `\$12\catcode
  `\&12\catcode `\#12\catcode `\^12\catcode `\_12\catcode `\%12\relax}%
\providecommand \@@startlink[1]{}%
\providecommand \@@endlink[0]{}%
\providecommand \url  [0]{\begingroup\@sanitize@url \@url }%
\providecommand \@url [1]{\endgroup\@href {#1}{\urlprefix }}%
\providecommand \urlprefix  [0]{URL }%
\providecommand \Eprint [0]{\href }%
\providecommand \doibase [0]{http://dx.doi.org/}%
\providecommand \selectlanguage [0]{\@gobble}%
\providecommand \bibinfo  [0]{\@secondoftwo}%
\providecommand \bibfield  [0]{\@secondoftwo}%
\providecommand \translation [1]{[#1]}%
\providecommand \BibitemOpen [0]{}%
\providecommand \bibitemStop [0]{}%
\providecommand \bibitemNoStop [0]{.\EOS\space}%
\providecommand \EOS [0]{\spacefactor3000\relax}%
\providecommand \BibitemShut  [1]{\csname bibitem#1\endcsname}%
\let\auto@bib@innerbib\@empty
\bibitem [{\citenamefont {Miracle}\ and\ \citenamefont
  {Senkov}(2017)}]{miracle_2017}%
  \BibitemOpen
  \bibfield  {author} {\bibinfo {author} {\bibfnamefont {D.}~\bibnamefont
  {Miracle}}\ and\ \bibinfo {author} {\bibfnamefont {O.}~\bibnamefont
  {Senkov}},\ }\href {\doibase https://doi.org/10.1016/j.actamat.2016.08.081}
  {\bibfield  {journal} {\bibinfo  {journal} {Acta Materialia}\ }\textbf
  {\bibinfo {volume} {122}},\ \bibinfo {pages} {448 } (\bibinfo {year}
  {2017})}\BibitemShut {NoStop}%
\bibitem [{\citenamefont {Rost}\ \emph {et~al.}(2015)\citenamefont {Rost},
  \citenamefont {Sachet}, \citenamefont {Borman}, \citenamefont {Moballegh},
  \citenamefont {Dickey}, \citenamefont {Hou}, \citenamefont {Jones},
  \citenamefont {Curtarolo},\ and\ \citenamefont {Maria}}]{rost_2015_ncomm}%
  \BibitemOpen
  \bibfield  {author} {\bibinfo {author} {\bibfnamefont {C.~M.}\ \bibnamefont
  {Rost}}, \bibinfo {author} {\bibfnamefont {E.}~\bibnamefont {Sachet}},
  \bibinfo {author} {\bibfnamefont {T.}~\bibnamefont {Borman}}, \bibinfo
  {author} {\bibfnamefont {A.}~\bibnamefont {Moballegh}}, \bibinfo {author}
  {\bibfnamefont {E.~C.}\ \bibnamefont {Dickey}}, \bibinfo {author}
  {\bibfnamefont {D.}~\bibnamefont {Hou}}, \bibinfo {author} {\bibfnamefont
  {J.~L.}\ \bibnamefont {Jones}}, \bibinfo {author} {\bibfnamefont
  {S.}~\bibnamefont {Curtarolo}}, \ and\ \bibinfo {author} {\bibfnamefont
  {J.-P.}\ \bibnamefont {Maria}},\ }\href {\doibase 10.1038/ncomms9485}
  {\bibfield  {journal} {\bibinfo  {journal} {Nature Communications}\ }\textbf
  {\bibinfo {volume} {6}},\ \bibinfo {pages} {8485} (\bibinfo {year}
  {2015})}\BibitemShut {NoStop}%
\bibitem [{\citenamefont {Sarkar}\ \emph {et~al.}(2019)\citenamefont {Sarkar},
  \citenamefont {Wang}, \citenamefont {Schiele}, \citenamefont {Chellali},
  \citenamefont {Bhattacharya}, \citenamefont {Wang}, \citenamefont
  {Brezesinski}, \citenamefont {Hahn}, \citenamefont {Velasco},\ and\
  \citenamefont {Breitung}}]{sarkar_2019_advmat}%
  \BibitemOpen
  \bibfield  {author} {\bibinfo {author} {\bibfnamefont {A.}~\bibnamefont
  {Sarkar}}, \bibinfo {author} {\bibfnamefont {Q.}~\bibnamefont {Wang}},
  \bibinfo {author} {\bibfnamefont {A.}~\bibnamefont {Schiele}}, \bibinfo
  {author} {\bibfnamefont {M.~R.}\ \bibnamefont {Chellali}}, \bibinfo {author}
  {\bibfnamefont {S.~S.}\ \bibnamefont {Bhattacharya}}, \bibinfo {author}
  {\bibfnamefont {D.}~\bibnamefont {Wang}}, \bibinfo {author} {\bibfnamefont
  {T.}~\bibnamefont {Brezesinski}}, \bibinfo {author} {\bibfnamefont
  {H.}~\bibnamefont {Hahn}}, \bibinfo {author} {\bibfnamefont {L.}~\bibnamefont
  {Velasco}}, \ and\ \bibinfo {author} {\bibfnamefont {B.}~\bibnamefont
  {Breitung}},\ }\href {\doibase 10.1002/adma.201806236} {\bibfield  {journal}
  {\bibinfo  {journal} {Advanced Materials}\ }\textbf {\bibinfo {volume}
  {31}},\ \bibinfo {pages} {1806236} (\bibinfo {year} {2019})}\BibitemShut
  {NoStop}%
\bibitem [{\citenamefont {Bérardan}\ \emph
  {et~al.}(2016{\natexlab{a}})\citenamefont {Bérardan}, \citenamefont
  {Franger}, \citenamefont {Dragoe}, \citenamefont {Meena},\ and\ \citenamefont
  {Dragoe}}]{berardan_2016_pss}%
  \BibitemOpen
  \bibfield  {author} {\bibinfo {author} {\bibfnamefont {D.}~\bibnamefont
  {Bérardan}}, \bibinfo {author} {\bibfnamefont {S.}~\bibnamefont {Franger}},
  \bibinfo {author} {\bibfnamefont {D.}~\bibnamefont {Dragoe}}, \bibinfo
  {author} {\bibfnamefont {A.~K.}\ \bibnamefont {Meena}}, \ and\ \bibinfo
  {author} {\bibfnamefont {N.}~\bibnamefont {Dragoe}},\ }\href {\doibase
  10.1002/pssr.201600043} {\bibfield  {journal} {\bibinfo  {journal} {physica
  status solidi (RRL) – Rapid Research Letters}\ }\textbf {\bibinfo {volume}
  {10}},\ \bibinfo {pages} {328} (\bibinfo {year}
  {2016}{\natexlab{a}})}\BibitemShut {NoStop}%
\bibitem [{\citenamefont {Bérardan}\ \emph
  {et~al.}(2016{\natexlab{b}})\citenamefont {Bérardan}, \citenamefont
  {Franger}, \citenamefont {Meena},\ and\ \citenamefont
  {Dragoe}}]{berardan_2016_jmca}%
  \BibitemOpen
  \bibfield  {author} {\bibinfo {author} {\bibfnamefont {D.}~\bibnamefont
  {Bérardan}}, \bibinfo {author} {\bibfnamefont {S.}~\bibnamefont {Franger}},
  \bibinfo {author} {\bibfnamefont {A.~K.}\ \bibnamefont {Meena}}, \ and\
  \bibinfo {author} {\bibfnamefont {N.}~\bibnamefont {Dragoe}},\ }\href
  {\doibase 10.1039/C6TA03249D} {\bibfield  {journal} {\bibinfo  {journal} {J.
  Mater. Chem. A}\ }\textbf {\bibinfo {volume} {4}},\ \bibinfo {pages} {9536}
  (\bibinfo {year} {2016}{\natexlab{b}})}\BibitemShut {NoStop}%
\bibitem [{\citenamefont {Braun}\ \emph {et~al.}(2018)\citenamefont {Braun},
  \citenamefont {Rost}, \citenamefont {Lim}, \citenamefont {Giri},
  \citenamefont {Olson}, \citenamefont {Kotsonis}, \citenamefont {Stan},
  \citenamefont {Brenner}, \citenamefont {Maria},\ and\ \citenamefont
  {Hopkins}}]{braun_2018_advmat}%
  \BibitemOpen
  \bibfield  {author} {\bibinfo {author} {\bibfnamefont {J.~L.}\ \bibnamefont
  {Braun}}, \bibinfo {author} {\bibfnamefont {C.~M.}\ \bibnamefont {Rost}},
  \bibinfo {author} {\bibfnamefont {M.}~\bibnamefont {Lim}}, \bibinfo {author}
  {\bibfnamefont {A.}~\bibnamefont {Giri}}, \bibinfo {author} {\bibfnamefont
  {D.~H.}\ \bibnamefont {Olson}}, \bibinfo {author} {\bibfnamefont {G.~N.}\
  \bibnamefont {Kotsonis}}, \bibinfo {author} {\bibfnamefont {G.}~\bibnamefont
  {Stan}}, \bibinfo {author} {\bibfnamefont {D.~W.}\ \bibnamefont {Brenner}},
  \bibinfo {author} {\bibfnamefont {J.-P.}\ \bibnamefont {Maria}}, \ and\
  \bibinfo {author} {\bibfnamefont {P.~E.}\ \bibnamefont {Hopkins}},\ }\href
  {\doibase 10.1002/adma.201805004} {\bibfield  {journal} {\bibinfo  {journal}
  {Advanced Materials}\ }\textbf {\bibinfo {volume} {30}},\ \bibinfo {pages}
  {1805004} (\bibinfo {year} {2018})}\BibitemShut {NoStop}%
\bibitem [{\citenamefont {Sarkar}\ \emph {et~al.}(2018)\citenamefont {Sarkar},
  \citenamefont {Djenadic}, \citenamefont {Wang}, \citenamefont {Hein},
  \citenamefont {Kautenburger}, \citenamefont {Clemens},\ and\ \citenamefont
  {Hahn}}]{sarkar_2018_jecs}%
  \BibitemOpen
  \bibfield  {author} {\bibinfo {author} {\bibfnamefont {A.}~\bibnamefont
  {Sarkar}}, \bibinfo {author} {\bibfnamefont {R.}~\bibnamefont {Djenadic}},
  \bibinfo {author} {\bibfnamefont {D.}~\bibnamefont {Wang}}, \bibinfo {author}
  {\bibfnamefont {C.}~\bibnamefont {Hein}}, \bibinfo {author} {\bibfnamefont
  {R.}~\bibnamefont {Kautenburger}}, \bibinfo {author} {\bibfnamefont
  {O.}~\bibnamefont {Clemens}}, \ and\ \bibinfo {author} {\bibfnamefont
  {H.}~\bibnamefont {Hahn}},\ }\href {\doibase
  https://doi.org/10.1016/j.jeurceramsoc.2017.12.058} {\bibfield  {journal}
  {\bibinfo  {journal} {Journal of the European Ceramic Society}\ }\textbf
  {\bibinfo {volume} {38}},\ \bibinfo {pages} {2318 } (\bibinfo {year}
  {2018})}\BibitemShut {NoStop}%
\bibitem [{\citenamefont {Jiang}\ \emph {et~al.}(2018)\citenamefont {Jiang},
  \citenamefont {Hu}, \citenamefont {Gild}, \citenamefont {Zhou}, \citenamefont
  {Nie}, \citenamefont {Qin}, \citenamefont {Harrington}, \citenamefont
  {Vecchio},\ and\ \citenamefont {Luo}}]{jiang_2018_sm}%
  \BibitemOpen
  \bibfield  {author} {\bibinfo {author} {\bibfnamefont {S.}~\bibnamefont
  {Jiang}}, \bibinfo {author} {\bibfnamefont {T.}~\bibnamefont {Hu}}, \bibinfo
  {author} {\bibfnamefont {J.}~\bibnamefont {Gild}}, \bibinfo {author}
  {\bibfnamefont {N.}~\bibnamefont {Zhou}}, \bibinfo {author} {\bibfnamefont
  {J.}~\bibnamefont {Nie}}, \bibinfo {author} {\bibfnamefont {M.}~\bibnamefont
  {Qin}}, \bibinfo {author} {\bibfnamefont {T.}~\bibnamefont {Harrington}},
  \bibinfo {author} {\bibfnamefont {K.}~\bibnamefont {Vecchio}}, \ and\
  \bibinfo {author} {\bibfnamefont {J.}~\bibnamefont {Luo}},\ }\href {\doibase
  https://doi.org/10.1016/j.scriptamat.2017.08.040} {\bibfield  {journal}
  {\bibinfo  {journal} {Scripta Materialia}\ }\textbf {\bibinfo {volume}
  {142}},\ \bibinfo {pages} {116 } (\bibinfo {year} {2018})}\BibitemShut
  {NoStop}%
\bibitem [{\citenamefont {Sharma}\ \emph {et~al.}(2018)\citenamefont {Sharma},
  \citenamefont {Musico}, \citenamefont {Gao}, \citenamefont {Hua},
  \citenamefont {May}, \citenamefont {Herklotz}, \citenamefont {Rastogi},
  \citenamefont {Mandrus}, \citenamefont {Yan}, \citenamefont {Lee},
  \citenamefont {Chisholm}, \citenamefont {Keppens},\ and\ \citenamefont
  {Ward}}]{sharma_2018_prbr}%
  \BibitemOpen
  \bibfield  {author} {\bibinfo {author} {\bibfnamefont {Y.}~\bibnamefont
  {Sharma}}, \bibinfo {author} {\bibfnamefont {B.~L.}\ \bibnamefont {Musico}},
  \bibinfo {author} {\bibfnamefont {X.}~\bibnamefont {Gao}}, \bibinfo {author}
  {\bibfnamefont {C.}~\bibnamefont {Hua}}, \bibinfo {author} {\bibfnamefont
  {A.~F.}\ \bibnamefont {May}}, \bibinfo {author} {\bibfnamefont
  {A.}~\bibnamefont {Herklotz}}, \bibinfo {author} {\bibfnamefont
  {A.}~\bibnamefont {Rastogi}}, \bibinfo {author} {\bibfnamefont
  {D.}~\bibnamefont {Mandrus}}, \bibinfo {author} {\bibfnamefont
  {J.}~\bibnamefont {Yan}}, \bibinfo {author} {\bibfnamefont {H.~N.}\
  \bibnamefont {Lee}}, \bibinfo {author} {\bibfnamefont {M.~F.}\ \bibnamefont
  {Chisholm}}, \bibinfo {author} {\bibfnamefont {V.}~\bibnamefont {Keppens}}, \
  and\ \bibinfo {author} {\bibfnamefont {T.~Z.}\ \bibnamefont {Ward}},\ }\href
  {\doibase 10.1103/PhysRevMaterials.2.060404} {\bibfield  {journal} {\bibinfo
  {journal} {Phys. Rev. Materials}\ }\textbf {\bibinfo {volume} {2}},\ \bibinfo
  {pages} {060404} (\bibinfo {year} {2018})}\BibitemShut {NoStop}%
\bibitem [{\citenamefont {Gild}\ \emph {et~al.}(2018)\citenamefont {Gild},
  \citenamefont {Samiee}, \citenamefont {Braun}, \citenamefont {Harrington},
  \citenamefont {Vega}, \citenamefont {Hopkins}, \citenamefont {Vecchio},\ and\
  \citenamefont {Luo}}]{gild_2018_jecs}%
  \BibitemOpen
  \bibfield  {author} {\bibinfo {author} {\bibfnamefont {J.}~\bibnamefont
  {Gild}}, \bibinfo {author} {\bibfnamefont {M.}~\bibnamefont {Samiee}},
  \bibinfo {author} {\bibfnamefont {J.~L.}\ \bibnamefont {Braun}}, \bibinfo
  {author} {\bibfnamefont {T.}~\bibnamefont {Harrington}}, \bibinfo {author}
  {\bibfnamefont {H.}~\bibnamefont {Vega}}, \bibinfo {author} {\bibfnamefont
  {P.~E.}\ \bibnamefont {Hopkins}}, \bibinfo {author} {\bibfnamefont
  {K.}~\bibnamefont {Vecchio}}, \ and\ \bibinfo {author} {\bibfnamefont
  {J.}~\bibnamefont {Luo}},\ }\href {\doibase
  https://doi.org/10.1016/j.jeurceramsoc.2018.04.010} {\bibfield  {journal}
  {\bibinfo  {journal} {Journal of the European Ceramic Society}\ }\textbf
  {\bibinfo {volume} {38}},\ \bibinfo {pages} {3578 } (\bibinfo {year}
  {2018})}\BibitemShut {NoStop}%
\bibitem [{\citenamefont {Chen}\ \emph {et~al.}(2018)\citenamefont {Chen},
  \citenamefont {Pei}, \citenamefont {Tang}, \citenamefont {Cheng},
  \citenamefont {Li}, \citenamefont {Li}, \citenamefont {Zhang},\ and\
  \citenamefont {An}}]{chen_2018_jecs}%
  \BibitemOpen
  \bibfield  {author} {\bibinfo {author} {\bibfnamefont {K.}~\bibnamefont
  {Chen}}, \bibinfo {author} {\bibfnamefont {X.}~\bibnamefont {Pei}}, \bibinfo
  {author} {\bibfnamefont {L.}~\bibnamefont {Tang}}, \bibinfo {author}
  {\bibfnamefont {H.}~\bibnamefont {Cheng}}, \bibinfo {author} {\bibfnamefont
  {Z.}~\bibnamefont {Li}}, \bibinfo {author} {\bibfnamefont {C.}~\bibnamefont
  {Li}}, \bibinfo {author} {\bibfnamefont {X.}~\bibnamefont {Zhang}}, \ and\
  \bibinfo {author} {\bibfnamefont {L.}~\bibnamefont {An}},\ }\href {\doibase
  https://doi.org/10.1016/j.jeurceramsoc.2018.04.063} {\bibfield  {journal}
  {\bibinfo  {journal} {Journal of the European Ceramic Society}\ }\textbf
  {\bibinfo {volume} {38}},\ \bibinfo {pages} {4161 } (\bibinfo {year}
  {2018})}\BibitemShut {NoStop}%
\bibitem [{\citenamefont {Dabrowa}\ \emph {et~al.}(2018)\citenamefont
  {Dabrowa}, \citenamefont {Stygar}, \citenamefont {Mikula}, \citenamefont
  {Knapik}, \citenamefont {Mroczka}, \citenamefont {Tejchman}, \citenamefont
  {Danielewski},\ and\ \citenamefont {Martin}}]{dabrowa_2018_ml}%
  \BibitemOpen
  \bibfield  {author} {\bibinfo {author} {\bibfnamefont {J.}~\bibnamefont
  {Dabrowa}}, \bibinfo {author} {\bibfnamefont {M.}~\bibnamefont {Stygar}},
  \bibinfo {author} {\bibfnamefont {A.}~\bibnamefont {Mikula}}, \bibinfo
  {author} {\bibfnamefont {A.}~\bibnamefont {Knapik}}, \bibinfo {author}
  {\bibfnamefont {K.}~\bibnamefont {Mroczka}}, \bibinfo {author} {\bibfnamefont
  {W.}~\bibnamefont {Tejchman}}, \bibinfo {author} {\bibfnamefont
  {M.}~\bibnamefont {Danielewski}}, \ and\ \bibinfo {author} {\bibfnamefont
  {M.}~\bibnamefont {Martin}},\ }\href {\doibase
  https://doi.org/10.1016/j.matlet.2017.12.148} {\bibfield  {journal} {\bibinfo
   {journal} {Materials Letters}\ }\textbf {\bibinfo {volume} {216}},\ \bibinfo
  {pages} {32 } (\bibinfo {year} {2018})}\BibitemShut {NoStop}%
\bibitem [{\citenamefont {Castle}\ \emph {et~al.}(2018)\citenamefont {Castle},
  \citenamefont {Csan{\~A}{\textexclamdown}di}, \citenamefont {Grasso},
  \citenamefont {Dusza},\ and\ \citenamefont {Reece}}]{castle_2018_srep}%
  \BibitemOpen
  \bibfield  {author} {\bibinfo {author} {\bibfnamefont {E.}~\bibnamefont
  {Castle}}, \bibinfo {author} {\bibfnamefont {T.}~\bibnamefont
  {Csan{\~A}{\textexclamdown}di}}, \bibinfo {author} {\bibfnamefont
  {S.}~\bibnamefont {Grasso}}, \bibinfo {author} {\bibfnamefont
  {J.}~\bibnamefont {Dusza}}, \ and\ \bibinfo {author} {\bibfnamefont
  {M.}~\bibnamefont {Reece}},\ }\href {\doibase 10.1038/s41598-018-26827-1}
  {\bibfield  {journal} {\bibinfo  {journal} {Scientific Reports}\ }\textbf
  {\bibinfo {volume} {8}},\ \bibinfo {pages} {8609} (\bibinfo {year}
  {2018})}\BibitemShut {NoStop}%
\bibitem [{\citenamefont {Yan}\ \emph {et~al.}(2018)\citenamefont {Yan},
  \citenamefont {Constantin}, \citenamefont {Lu}, \citenamefont {Silvain},
  \citenamefont {Nastasi},\ and\ \citenamefont {Cui}}]{yan_2018_jacs}%
  \BibitemOpen
  \bibfield  {author} {\bibinfo {author} {\bibfnamefont {X.}~\bibnamefont
  {Yan}}, \bibinfo {author} {\bibfnamefont {L.}~\bibnamefont {Constantin}},
  \bibinfo {author} {\bibfnamefont {Y.}~\bibnamefont {Lu}}, \bibinfo {author}
  {\bibfnamefont {J.-F.}\ \bibnamefont {Silvain}}, \bibinfo {author}
  {\bibfnamefont {M.}~\bibnamefont {Nastasi}}, \ and\ \bibinfo {author}
  {\bibfnamefont {B.}~\bibnamefont {Cui}},\ }\href {\doibase
  10.1111/jace.15779} {\bibfield  {journal} {\bibinfo  {journal} {Journal of
  the American Ceramic Society}\ }\textbf {\bibinfo {volume} {101}},\ \bibinfo
  {pages} {4486} (\bibinfo {year} {2018})}\BibitemShut {NoStop}%
\bibitem [{\citenamefont {Gild}\ \emph {et~al.}(2016)\citenamefont {Gild},
  \citenamefont {Zhang}, \citenamefont {Harrington}, \citenamefont {Jiang},
  \citenamefont {Hu}, \citenamefont {Quinn}, \citenamefont {Mellor},
  \citenamefont {Zhou}, \citenamefont {Vecchio},\ and\ \citenamefont
  {Luo}}]{gild_2016_srep}%
  \BibitemOpen
  \bibfield  {author} {\bibinfo {author} {\bibfnamefont {J.}~\bibnamefont
  {Gild}}, \bibinfo {author} {\bibfnamefont {Y.}~\bibnamefont {Zhang}},
  \bibinfo {author} {\bibfnamefont {T.}~\bibnamefont {Harrington}}, \bibinfo
  {author} {\bibfnamefont {S.}~\bibnamefont {Jiang}}, \bibinfo {author}
  {\bibfnamefont {T.}~\bibnamefont {Hu}}, \bibinfo {author} {\bibfnamefont
  {M.~C.}\ \bibnamefont {Quinn}}, \bibinfo {author} {\bibfnamefont {W.~M.}\
  \bibnamefont {Mellor}}, \bibinfo {author} {\bibfnamefont {N.}~\bibnamefont
  {Zhou}}, \bibinfo {author} {\bibfnamefont {K.}~\bibnamefont {Vecchio}}, \
  and\ \bibinfo {author} {\bibfnamefont {J.}~\bibnamefont {Luo}},\ }\href
  {\doibase 10.1038/srep37946} {\bibfield  {journal} {\bibinfo  {journal}
  {Scientific Reports}\ }\textbf {\bibinfo {volume} {6}},\ \bibinfo {pages}
  {37946} (\bibinfo {year} {2016})}\BibitemShut {NoStop}%
\bibitem [{\citenamefont {Deng}\ \emph {et~al.}(2020)\citenamefont {Deng},
  \citenamefont {Olvera}, \citenamefont {Casamento}, \citenamefont {Lopez},
  \citenamefont {Williams}, \citenamefont {Lu}, \citenamefont {Shi},
  \citenamefont {Poudeu},\ and\ \citenamefont {Kioupakis}}]{deng_2020_chemmat}%
  \BibitemOpen
  \bibfield  {author} {\bibinfo {author} {\bibfnamefont {Z.}~\bibnamefont
  {Deng}}, \bibinfo {author} {\bibfnamefont {A.}~\bibnamefont {Olvera}},
  \bibinfo {author} {\bibfnamefont {J.}~\bibnamefont {Casamento}}, \bibinfo
  {author} {\bibfnamefont {J.~S.}\ \bibnamefont {Lopez}}, \bibinfo {author}
  {\bibfnamefont {L.}~\bibnamefont {Williams}}, \bibinfo {author}
  {\bibfnamefont {R.}~\bibnamefont {Lu}}, \bibinfo {author} {\bibfnamefont
  {G.}~\bibnamefont {Shi}}, \bibinfo {author} {\bibfnamefont {P.~F.~P.}\
  \bibnamefont {Poudeu}}, \ and\ \bibinfo {author} {\bibfnamefont
  {E.}~\bibnamefont {Kioupakis}},\ }\href {\doibase
  10.1021/acs.chemmater.0c01555} {\bibfield  {journal} {\bibinfo  {journal}
  {Chemistry of Materials}\ }\textbf {\bibinfo {volume} {32}},\ \bibinfo
  {pages} {6070} (\bibinfo {year} {2020})}\BibitemShut {NoStop}%
\bibitem [{\citenamefont {Meisenheimer}\ \emph {et~al.}(2017)\citenamefont
  {Meisenheimer}, \citenamefont {Kratofil},\ and\ \citenamefont
  {Heron}}]{meisenheimer_2017_srep}%
  \BibitemOpen
  \bibfield  {author} {\bibinfo {author} {\bibfnamefont {P.~B.}\ \bibnamefont
  {Meisenheimer}}, \bibinfo {author} {\bibfnamefont {T.~J.}\ \bibnamefont
  {Kratofil}}, \ and\ \bibinfo {author} {\bibfnamefont {J.~T.}\ \bibnamefont
  {Heron}},\ }\href {\doibase 10.1038/s41598-017-13810-5} {\bibfield  {journal}
  {\bibinfo  {journal} {Scientific Reports}\ }\textbf {\bibinfo {volume} {7}},\
  \bibinfo {pages} {13344} (\bibinfo {year} {2017})}\BibitemShut {NoStop}%
\bibitem [{\citenamefont {Zhang}\ \emph {et~al.}(2019)\citenamefont {Zhang},
  \citenamefont {Yan}, \citenamefont {Calder}, \citenamefont {Zheng},
  \citenamefont {McGuire}, \citenamefont {Abernathy}, \citenamefont {Ren},
  \citenamefont {Lapidus}, \citenamefont {Page}, \citenamefont {Zheng},
  \citenamefont {Freeland}, \citenamefont {Budai},\ and\ \citenamefont
  {Hermann}}]{zhang_2019_chemmat}%
  \BibitemOpen
  \bibfield  {author} {\bibinfo {author} {\bibfnamefont {J.}~\bibnamefont
  {Zhang}}, \bibinfo {author} {\bibfnamefont {J.}~\bibnamefont {Yan}}, \bibinfo
  {author} {\bibfnamefont {S.}~\bibnamefont {Calder}}, \bibinfo {author}
  {\bibfnamefont {Q.}~\bibnamefont {Zheng}}, \bibinfo {author} {\bibfnamefont
  {M.~A.}\ \bibnamefont {McGuire}}, \bibinfo {author} {\bibfnamefont {D.~L.}\
  \bibnamefont {Abernathy}}, \bibinfo {author} {\bibfnamefont {Y.}~\bibnamefont
  {Ren}}, \bibinfo {author} {\bibfnamefont {S.~H.}\ \bibnamefont {Lapidus}},
  \bibinfo {author} {\bibfnamefont {K.}~\bibnamefont {Page}}, \bibinfo {author}
  {\bibfnamefont {H.}~\bibnamefont {Zheng}}, \bibinfo {author} {\bibfnamefont
  {J.~W.}\ \bibnamefont {Freeland}}, \bibinfo {author} {\bibfnamefont {J.~D.}\
  \bibnamefont {Budai}}, \ and\ \bibinfo {author} {\bibfnamefont {R.~P.}\
  \bibnamefont {Hermann}},\ }\href {\doibase 10.1021/acs.chemmater.9b00624}
  {\bibfield  {journal} {\bibinfo  {journal} {Chemistry of Materials}\ }\textbf
  {\bibinfo {volume} {31}},\ \bibinfo {pages} {3705} (\bibinfo {year}
  {2019})}\BibitemShut {NoStop}%
\bibitem [{\citenamefont {Jimenez-Segura}\ \emph {et~al.}(2019)\citenamefont
  {Jimenez-Segura}, \citenamefont {Takayama}, \citenamefont {Bérardan},
  \citenamefont {Hoser}, \citenamefont {Reehuis}, \citenamefont {Takagi},\ and\
  \citenamefont {Dragoe}}]{jimenez_2019_apl}%
  \BibitemOpen
  \bibfield  {author} {\bibinfo {author} {\bibfnamefont {M.~P.}\ \bibnamefont
  {Jimenez-Segura}}, \bibinfo {author} {\bibfnamefont {T.}~\bibnamefont
  {Takayama}}, \bibinfo {author} {\bibfnamefont {D.}~\bibnamefont {Bérardan}},
  \bibinfo {author} {\bibfnamefont {A.}~\bibnamefont {Hoser}}, \bibinfo
  {author} {\bibfnamefont {M.}~\bibnamefont {Reehuis}}, \bibinfo {author}
  {\bibfnamefont {H.}~\bibnamefont {Takagi}}, \ and\ \bibinfo {author}
  {\bibfnamefont {N.}~\bibnamefont {Dragoe}},\ }\href {\doibase
  10.1063/1.5091787} {\bibfield  {journal} {\bibinfo  {journal} {Applied
  Physics Letters}\ }\textbf {\bibinfo {volume} {114}},\ \bibinfo {pages}
  {122401} (\bibinfo {year} {2019})}\BibitemShut {NoStop}%
\bibitem [{\citenamefont {Shull}\ \emph {et~al.}(1951)\citenamefont {Shull},
  \citenamefont {Strauser},\ and\ \citenamefont {Wollan}}]{shull_1951}%
  \BibitemOpen
  \bibfield  {author} {\bibinfo {author} {\bibfnamefont {C.~G.}\ \bibnamefont
  {Shull}}, \bibinfo {author} {\bibfnamefont {W.~A.}\ \bibnamefont {Strauser}},
  \ and\ \bibinfo {author} {\bibfnamefont {E.~O.}\ \bibnamefont {Wollan}},\
  }\href {\doibase 10.1103/PhysRev.83.333} {\bibfield  {journal} {\bibinfo
  {journal} {Phys. Rev.}\ }\textbf {\bibinfo {volume} {83}},\ \bibinfo {pages}
  {333} (\bibinfo {year} {1951})}\BibitemShut {NoStop}%
\bibitem [{\citenamefont {Roth}(1958)}]{roth_1958}%
  \BibitemOpen
  \bibfield  {author} {\bibinfo {author} {\bibfnamefont {W.~L.}\ \bibnamefont
  {Roth}},\ }\href {\doibase 10.1103/PhysRev.110.1333} {\bibfield  {journal}
  {\bibinfo  {journal} {Phys. Rev.}\ }\textbf {\bibinfo {volume} {110}},\
  \bibinfo {pages} {1333} (\bibinfo {year} {1958})}\BibitemShut {NoStop}%
\bibitem [{\citenamefont {Frandsen}\ \emph {et~al.}(2020)\citenamefont
  {Frandsen}, \citenamefont {Petersen}, \citenamefont {Ducharme}, \citenamefont
  {Shaw}, \citenamefont {Gibson}, \citenamefont {Winn}, \citenamefont {Yan},
  \citenamefont {Zhang}, \citenamefont {Manley},\ and\ \citenamefont
  {Hermann}}]{frandsen_2020_prm}%
  \BibitemOpen
  \bibfield  {author} {\bibinfo {author} {\bibfnamefont {B.~A.}\ \bibnamefont
  {Frandsen}}, \bibinfo {author} {\bibfnamefont {K.~A.}\ \bibnamefont
  {Petersen}}, \bibinfo {author} {\bibfnamefont {N.~A.}\ \bibnamefont
  {Ducharme}}, \bibinfo {author} {\bibfnamefont {A.~G.}\ \bibnamefont {Shaw}},
  \bibinfo {author} {\bibfnamefont {E.~J.}\ \bibnamefont {Gibson}}, \bibinfo
  {author} {\bibfnamefont {B.}~\bibnamefont {Winn}}, \bibinfo {author}
  {\bibfnamefont {J.}~\bibnamefont {Yan}}, \bibinfo {author} {\bibfnamefont
  {J.}~\bibnamefont {Zhang}}, \bibinfo {author} {\bibfnamefont {M.~E.}\
  \bibnamefont {Manley}}, \ and\ \bibinfo {author} {\bibfnamefont {R.~P.}\
  \bibnamefont {Hermann}},\ }\href {\doibase 10.1103/PhysRevMaterials.4.074405}
  {\bibfield  {journal} {\bibinfo  {journal} {Phys. Rev. Materials}\ }\textbf
  {\bibinfo {volume} {4}},\ \bibinfo {pages} {074405} (\bibinfo {year}
  {2020})}\BibitemShut {NoStop}%
\bibitem [{\citenamefont {Menshikov}\ \emph {et~al.}(1991)\citenamefont
  {Menshikov}, \citenamefont {Dorofeev}, \citenamefont {Klimenko},\ and\
  \citenamefont {Mironova}}]{menshikov_1991}%
  \BibitemOpen
  \bibfield  {author} {\bibinfo {author} {\bibfnamefont {A.~Z.}\ \bibnamefont
  {Menshikov}}, \bibinfo {author} {\bibfnamefont {Y.~A.}\ \bibnamefont
  {Dorofeev}}, \bibinfo {author} {\bibfnamefont {A.~G.}\ \bibnamefont
  {Klimenko}}, \ and\ \bibinfo {author} {\bibfnamefont {N.~A.}\ \bibnamefont
  {Mironova}},\ }\href {\doibase 10.1002/pssb.2221640130} {\bibfield  {journal}
  {\bibinfo  {journal} {physica status solidi (b)}\ }\textbf {\bibinfo {volume}
  {164}},\ \bibinfo {pages} {275} (\bibinfo {year} {1991})}\BibitemShut
  {NoStop}%
\bibitem [{\citenamefont {Rodic}\ \emph {et~al.}(2000)\citenamefont {Rodic},
  \citenamefont {Spasojevic}, \citenamefont {Kusigerski}, \citenamefont
  {Tellgren},\ and\ \citenamefont {Rundlof}}]{rodic_2000}%
  \BibitemOpen
  \bibfield  {author} {\bibinfo {author} {\bibfnamefont {D.}~\bibnamefont
  {Rodic}}, \bibinfo {author} {\bibfnamefont {V.}~\bibnamefont {Spasojevic}},
  \bibinfo {author} {\bibfnamefont {V.}~\bibnamefont {Kusigerski}}, \bibinfo
  {author} {\bibfnamefont {R.}~\bibnamefont {Tellgren}}, \ and\ \bibinfo
  {author} {\bibfnamefont {H.}~\bibnamefont {Rundlof}},\ }\href {\doibase
  10.1002/1521-3951(200004)218:2<527::AID-PSSB527>3.0.CO;2-I} {\bibfield
  {journal} {\bibinfo  {journal} {physica status solidi (b)}\ }\textbf
  {\bibinfo {volume} {218}},\ \bibinfo {pages} {527} (\bibinfo {year}
  {2000})}\BibitemShut {NoStop}%
\bibitem [{\citenamefont {Seehra}\ \emph {et~al.}(1988)\citenamefont {Seehra},
  \citenamefont {Dean},\ and\ \citenamefont {Kannan}}]{seehra_1988}%
  \BibitemOpen
  \bibfield  {author} {\bibinfo {author} {\bibfnamefont {M.~S.}\ \bibnamefont
  {Seehra}}, \bibinfo {author} {\bibfnamefont {J.~C.}\ \bibnamefont {Dean}}, \
  and\ \bibinfo {author} {\bibfnamefont {R.}~\bibnamefont {Kannan}},\ }\href
  {\doibase 10.1103/PhysRevB.37.5864} {\bibfield  {journal} {\bibinfo
  {journal} {Phys. Rev. B}\ }\textbf {\bibinfo {volume} {37}},\ \bibinfo
  {pages} {5864} (\bibinfo {year} {1988})}\BibitemShut {NoStop}%
\bibitem [{\citenamefont {Rost}\ \emph {et~al.}(2017)\citenamefont {Rost},
  \citenamefont {Rak}, \citenamefont {Brenner},\ and\ \citenamefont
  {Maria}}]{rost_2017_jacs}%
  \BibitemOpen
  \bibfield  {author} {\bibinfo {author} {\bibfnamefont {C.~M.}\ \bibnamefont
  {Rost}}, \bibinfo {author} {\bibfnamefont {Z.}~\bibnamefont {Rak}}, \bibinfo
  {author} {\bibfnamefont {D.~W.}\ \bibnamefont {Brenner}}, \ and\ \bibinfo
  {author} {\bibfnamefont {J.-P.}\ \bibnamefont {Maria}},\ }\href {\doibase
  10.1111/jace.14756} {\bibfield  {journal} {\bibinfo  {journal} {Journal of
  the American Ceramic Society}\ }\textbf {\bibinfo {volume} {100}},\ \bibinfo
  {pages} {2732} (\bibinfo {year} {2017})}\BibitemShut {NoStop}%
\bibitem [{\citenamefont {Rak}\ \emph {et~al.}(2016)\citenamefont {Rak},
  \citenamefont {Rost}, \citenamefont {Lim}, \citenamefont {Sarker},
  \citenamefont {Toher}, \citenamefont {Curtarolo}, \citenamefont {Maria},\
  and\ \citenamefont {Brenner}}]{rak_2016_jap}%
  \BibitemOpen
  \bibfield  {author} {\bibinfo {author} {\bibfnamefont {Z.}~\bibnamefont
  {Rak}}, \bibinfo {author} {\bibfnamefont {C.~M.}\ \bibnamefont {Rost}},
  \bibinfo {author} {\bibfnamefont {M.}~\bibnamefont {Lim}}, \bibinfo {author}
  {\bibfnamefont {P.}~\bibnamefont {Sarker}}, \bibinfo {author} {\bibfnamefont
  {C.}~\bibnamefont {Toher}}, \bibinfo {author} {\bibfnamefont
  {S.}~\bibnamefont {Curtarolo}}, \bibinfo {author} {\bibfnamefont {J.-P.}\
  \bibnamefont {Maria}}, \ and\ \bibinfo {author} {\bibfnamefont {D.~W.}\
  \bibnamefont {Brenner}},\ }\href {\doibase 10.1063/1.4962135} {\bibfield
  {journal} {\bibinfo  {journal} {Journal of Applied Physics}\ }\textbf
  {\bibinfo {volume} {120}},\ \bibinfo {pages} {095105} (\bibinfo {year}
  {2016})}\BibitemShut {NoStop}%
\bibitem [{\citenamefont {Goodenough}(1958)}]{goodenough_1958}%
  \BibitemOpen
  \bibfield  {author} {\bibinfo {author} {\bibfnamefont {J.~B.}\ \bibnamefont
  {Goodenough}},\ }\href {\doibase
  https://doi.org/10.1016/0022-3697(58)90107-0} {\bibfield  {journal} {\bibinfo
   {journal} {Journal of Physics and Chemistry of Solids}\ }\textbf {\bibinfo
  {volume} {6}},\ \bibinfo {pages} {287 } (\bibinfo {year} {1958})}\BibitemShut
  {NoStop}%
\bibitem [{\citenamefont {Kanamori}(1959)}]{kanamori_1959}%
  \BibitemOpen
  \bibfield  {author} {\bibinfo {author} {\bibfnamefont {J.}~\bibnamefont
  {Kanamori}},\ }\href {\doibase https://doi.org/10.1016/0022-3697(59)90061-7}
  {\bibfield  {journal} {\bibinfo  {journal} {Journal of Physics and Chemistry
  of Solids}\ }\textbf {\bibinfo {volume} {10}},\ \bibinfo {pages} {87 }
  (\bibinfo {year} {1959})}\BibitemShut {NoStop}%
\bibitem [{\citenamefont {Deng}\ \emph {et~al.}(2010)\citenamefont {Deng},
  \citenamefont {Li}, \citenamefont {Li}, \citenamefont {Xia}, \citenamefont
  {Walsh},\ and\ \citenamefont {Wei}}]{deng_apl_2010}%
  \BibitemOpen
  \bibfield  {author} {\bibinfo {author} {\bibfnamefont {H.-X.}\ \bibnamefont
  {Deng}}, \bibinfo {author} {\bibfnamefont {J.}~\bibnamefont {Li}}, \bibinfo
  {author} {\bibfnamefont {S.-S.}\ \bibnamefont {Li}}, \bibinfo {author}
  {\bibfnamefont {J.-B.}\ \bibnamefont {Xia}}, \bibinfo {author} {\bibfnamefont
  {A.}~\bibnamefont {Walsh}}, \ and\ \bibinfo {author} {\bibfnamefont {S.-H.}\
  \bibnamefont {Wei}},\ }\href {\doibase 10.1063/1.3402772} {\bibfield
  {journal} {\bibinfo  {journal} {Applied Physics Letters}\ }\textbf {\bibinfo
  {volume} {96}},\ \bibinfo {pages} {162508} (\bibinfo {year}
  {2010})}\BibitemShut {NoStop}%
\bibitem [{\citenamefont {K\"odderitzsch}\ \emph {et~al.}(2002)\citenamefont
  {K\"odderitzsch}, \citenamefont {Hergert}, \citenamefont {Temmerman},
  \citenamefont {Szotek}, \citenamefont {Ernst},\ and\ \citenamefont
  {Winter}}]{kodderitzsch_2002_prb}%
  \BibitemOpen
  \bibfield  {author} {\bibinfo {author} {\bibfnamefont {D.}~\bibnamefont
  {K\"odderitzsch}}, \bibinfo {author} {\bibfnamefont {W.}~\bibnamefont
  {Hergert}}, \bibinfo {author} {\bibfnamefont {W.~M.}\ \bibnamefont
  {Temmerman}}, \bibinfo {author} {\bibfnamefont {Z.}~\bibnamefont {Szotek}},
  \bibinfo {author} {\bibfnamefont {A.}~\bibnamefont {Ernst}}, \ and\ \bibinfo
  {author} {\bibfnamefont {H.}~\bibnamefont {Winter}},\ }\href {\doibase
  10.1103/PhysRevB.66.064434} {\bibfield  {journal} {\bibinfo  {journal} {Phys.
  Rev. B}\ }\textbf {\bibinfo {volume} {66}},\ \bibinfo {pages} {064434}
  (\bibinfo {year} {2002})}\BibitemShut {NoStop}%
\bibitem [{\citenamefont {R\'{a}k}\ and\ \citenamefont
  {Brenner}(2020)}]{rak_2020_jap}%
  \BibitemOpen
  \bibfield  {author} {\bibinfo {author} {\bibfnamefont {Z.}~\bibnamefont
  {R\'{a}k}}\ and\ \bibinfo {author} {\bibfnamefont {D.~W.}\ \bibnamefont
  {Brenner}},\ }\href {\doibase 10.1063/5.0008258} {\bibfield  {journal}
  {\bibinfo  {journal} {Journal of Applied Physics}\ }\textbf {\bibinfo
  {volume} {127}},\ \bibinfo {pages} {185108} (\bibinfo {year}
  {2020})}\BibitemShut {NoStop}%
\bibitem [{\citenamefont {Hutchings}\ and\ \citenamefont
  {Samuelsen}(1972)}]{hutchings_1972_prb}%
  \BibitemOpen
  \bibfield  {author} {\bibinfo {author} {\bibfnamefont {M.~T.}\ \bibnamefont
  {Hutchings}}\ and\ \bibinfo {author} {\bibfnamefont {E.~J.}\ \bibnamefont
  {Samuelsen}},\ }\href {\doibase 10.1103/PhysRevB.6.3447} {\bibfield
  {journal} {\bibinfo  {journal} {Phys. Rev. B}\ }\textbf {\bibinfo {volume}
  {6}},\ \bibinfo {pages} {3447} (\bibinfo {year} {1972})}\BibitemShut
  {NoStop}%
\bibitem [{\citenamefont {Tomiyasu}\ and\ \citenamefont
  {Itoh}(2006)}]{tomiyasu_2006_jpsj}%
  \BibitemOpen
  \bibfield  {author} {\bibinfo {author} {\bibfnamefont {K.}~\bibnamefont
  {Tomiyasu}}\ and\ \bibinfo {author} {\bibfnamefont {S.}~\bibnamefont
  {Itoh}},\ }\href {\doibase 10.1143/JPSJ.75.084708} {\bibfield  {journal}
  {\bibinfo  {journal} {Journal of the Physical Society of Japan}\ }\textbf
  {\bibinfo {volume} {75}},\ \bibinfo {pages} {084708} (\bibinfo {year}
  {2006})}\BibitemShut {NoStop}%
\bibitem [{\citenamefont {Yamamoto}\ and\ \citenamefont
  {Nagamiya}(1972)}]{yamamoto_1972_jpsj}%
  \BibitemOpen
  \bibfield  {author} {\bibinfo {author} {\bibfnamefont {Y.}~\bibnamefont
  {Yamamoto}}\ and\ \bibinfo {author} {\bibfnamefont {T.}~\bibnamefont
  {Nagamiya}},\ }\href {\doibase 10.1143/JPSJ.32.1248} {\bibfield  {journal}
  {\bibinfo  {journal} {Journal of the Physical Society of Japan}\ }\textbf
  {\bibinfo {volume} {32}},\ \bibinfo {pages} {1248} (\bibinfo {year}
  {1972})}\BibitemShut {NoStop}%
\bibitem [{\citenamefont {Fishman}\ \emph {et~al.}(2018)\citenamefont
  {Fishman}, \citenamefont {Fernandez-Baca},\ and\ \citenamefont
  {Room}}]{fishman_2018}%
  \BibitemOpen
  \bibfield  {author} {\bibinfo {author} {\bibfnamefont {R.}~\bibnamefont
  {Fishman}}, \bibinfo {author} {\bibfnamefont {J.}~\bibnamefont
  {Fernandez-Baca}}, \ and\ \bibinfo {author} {\bibfnamefont {T.}~\bibnamefont
  {Room}},\ }\href@noop {} {\emph {\bibinfo {title} {Spin-Wave Theory: and its
  Applications to Neutron Scattering and THz Spectroscopy}}}\ (\bibinfo
  {publisher} {Morgan \& Claypool Publishers},\ \bibinfo {address} {San
  Rafael},\ \bibinfo {year} {2018})\BibitemShut {NoStop}%
\bibitem [{\citenamefont {Berlijn}(2011)}]{berlijn_2011_thesis}%
  \BibitemOpen
  \bibfield  {author} {\bibinfo {author} {\bibfnamefont {T.}~\bibnamefont
  {Berlijn}},\ }\emph {\bibinfo {title} {Effects of Disordered Dopants on the
  Electronic Structure of Functional Materials: Wannier Function-Based First
  Principles Methods for Disordered Systems}},\ \href@noop {} {Ph.D. thesis},\
  \bibinfo  {school} {Stony Brook University} (\bibinfo {year}
  {2011})\BibitemShut {NoStop}%
\bibitem [{yas()}]{yaspinwave}%
  \BibitemOpen
  \href@noop {} {}\bibinfo {note}
  {\url{https://github.com/g1257/YaSpinWave}}\BibitemShut {NoStop}%
\bibitem [{\citenamefont {Buczek}\ \emph {et~al.}(2016)\citenamefont {Buczek},
  \citenamefont {Sandratskii}, \citenamefont {Buczek}, \citenamefont {Thomas},
  \citenamefont {Vignale},\ and\ \citenamefont {Ernst}}]{buczek_2016_prb}%
  \BibitemOpen
  \bibfield  {author} {\bibinfo {author} {\bibfnamefont {P.}~\bibnamefont
  {Buczek}}, \bibinfo {author} {\bibfnamefont {L.~M.}\ \bibnamefont
  {Sandratskii}}, \bibinfo {author} {\bibfnamefont {N.}~\bibnamefont {Buczek}},
  \bibinfo {author} {\bibfnamefont {S.}~\bibnamefont {Thomas}}, \bibinfo
  {author} {\bibfnamefont {G.}~\bibnamefont {Vignale}}, \ and\ \bibinfo
  {author} {\bibfnamefont {A.}~\bibnamefont {Ernst}},\ }\href {\doibase
  10.1103/PhysRevB.94.054407} {\bibfield  {journal} {\bibinfo  {journal} {Phys.
  Rev. B}\ }\textbf {\bibinfo {volume} {94}},\ \bibinfo {pages} {054407}
  (\bibinfo {year} {2016})}\BibitemShut {NoStop}%
\bibitem [{\citenamefont {Bai}\ \emph {et~al.}(2019)\citenamefont {Bai},
  \citenamefont {Paddison}, \citenamefont {Kapit}, \citenamefont {Koohpayeh},
  \citenamefont {Wen}, \citenamefont {Dutton}, \citenamefont {Savici},
  \citenamefont {Kolesnikov}, \citenamefont {Granroth}, \citenamefont
  {Broholm}, \citenamefont {Chalker},\ and\ \citenamefont
  {Mourigal}}]{bai_2019_prl}%
  \BibitemOpen
  \bibfield  {author} {\bibinfo {author} {\bibfnamefont {X.}~\bibnamefont
  {Bai}}, \bibinfo {author} {\bibfnamefont {J.~A.~M.}\ \bibnamefont
  {Paddison}}, \bibinfo {author} {\bibfnamefont {E.}~\bibnamefont {Kapit}},
  \bibinfo {author} {\bibfnamefont {S.~M.}\ \bibnamefont {Koohpayeh}}, \bibinfo
  {author} {\bibfnamefont {J.-J.}\ \bibnamefont {Wen}}, \bibinfo {author}
  {\bibfnamefont {S.~E.}\ \bibnamefont {Dutton}}, \bibinfo {author}
  {\bibfnamefont {A.~T.}\ \bibnamefont {Savici}}, \bibinfo {author}
  {\bibfnamefont {A.~I.}\ \bibnamefont {Kolesnikov}}, \bibinfo {author}
  {\bibfnamefont {G.~E.}\ \bibnamefont {Granroth}}, \bibinfo {author}
  {\bibfnamefont {C.~L.}\ \bibnamefont {Broholm}}, \bibinfo {author}
  {\bibfnamefont {J.~T.}\ \bibnamefont {Chalker}}, \ and\ \bibinfo {author}
  {\bibfnamefont {M.}~\bibnamefont {Mourigal}},\ }\href {\doibase
  10.1103/PhysRevLett.122.097201} {\bibfield  {journal} {\bibinfo  {journal}
  {Phys. Rev. Lett.}\ }\textbf {\bibinfo {volume} {122}},\ \bibinfo {pages}
  {097201} (\bibinfo {year} {2019})}\BibitemShut {NoStop}%
\bibitem [{\citenamefont {Buyers}\ \emph {et~al.}(1973)\citenamefont {Buyers},
  \citenamefont {Pepper},\ and\ \citenamefont {Elliott}}]{buyers_1973}%
  \BibitemOpen
  \bibfield  {author} {\bibinfo {author} {\bibfnamefont {W.~J.~L.}\
  \bibnamefont {Buyers}}, \bibinfo {author} {\bibfnamefont {D.~E.}\
  \bibnamefont {Pepper}}, \ and\ \bibinfo {author} {\bibfnamefont {R.~J.}\
  \bibnamefont {Elliott}},\ }\href {\doibase 10.1088/0022-3719/6/11/019}
  {\bibfield  {journal} {\bibinfo  {journal} {Journal of Physics C: Solid State
  Physics}\ }\textbf {\bibinfo {volume} {6}},\ \bibinfo {pages} {1933}
  (\bibinfo {year} {1973})}\BibitemShut {NoStop}%
\bibitem [{\citenamefont {Samarakoon}\ \emph {et~al.}(2017)\citenamefont
  {Samarakoon}, \citenamefont {Banerjee}, \citenamefont {Zhang}, \citenamefont
  {Kamiya}, \citenamefont {Nagler}, \citenamefont {Tennant}, \citenamefont
  {Lee},\ and\ \citenamefont {Batista}}]{samarakoon_2017_prb}%
  \BibitemOpen
  \bibfield  {author} {\bibinfo {author} {\bibfnamefont {A.~M.}\ \bibnamefont
  {Samarakoon}}, \bibinfo {author} {\bibfnamefont {A.}~\bibnamefont
  {Banerjee}}, \bibinfo {author} {\bibfnamefont {S.-S.}\ \bibnamefont {Zhang}},
  \bibinfo {author} {\bibfnamefont {Y.}~\bibnamefont {Kamiya}}, \bibinfo
  {author} {\bibfnamefont {S.~E.}\ \bibnamefont {Nagler}}, \bibinfo {author}
  {\bibfnamefont {D.~A.}\ \bibnamefont {Tennant}}, \bibinfo {author}
  {\bibfnamefont {S.-H.}\ \bibnamefont {Lee}}, \ and\ \bibinfo {author}
  {\bibfnamefont {C.~D.}\ \bibnamefont {Batista}},\ }\href {\doibase
  10.1103/PhysRevB.96.134408} {\bibfield  {journal} {\bibinfo  {journal} {Phys.
  Rev. B}\ }\textbf {\bibinfo {volume} {96}},\ \bibinfo {pages} {134408}
  (\bibinfo {year} {2017})}\BibitemShut {NoStop}%
\bibitem [{\citenamefont {Th\'ebaud}\ \emph {et~al.}(2020)\citenamefont
  {Th\'ebaud}, \citenamefont {Polanco}, \citenamefont {Lindsay},\ and\
  \citenamefont {Berlijn}}]{thebaud_2020}%
  \BibitemOpen
  \bibfield  {author} {\bibinfo {author} {\bibfnamefont {S.}~\bibnamefont
  {Th\'ebaud}}, \bibinfo {author} {\bibfnamefont {C.~A.}\ \bibnamefont
  {Polanco}}, \bibinfo {author} {\bibfnamefont {L.}~\bibnamefont {Lindsay}}, \
  and\ \bibinfo {author} {\bibfnamefont {T.}~\bibnamefont {Berlijn}},\ }\href
  {\doibase 10.1103/PhysRevB.102.094206} {\bibfield  {journal} {\bibinfo
  {journal} {Phys. Rev. B}\ }\textbf {\bibinfo {volume} {102}},\ \bibinfo
  {pages} {094206} (\bibinfo {year} {2020})}\BibitemShut {NoStop}%
\bibitem [{\citenamefont {Berardan}\ \emph {et~al.}(2017)\citenamefont
  {Berardan}, \citenamefont {Meena}, \citenamefont {Franger}, \citenamefont
  {Herrero},\ and\ \citenamefont {Dragoe}}]{berardan_2017_jac}%
  \BibitemOpen
  \bibfield  {author} {\bibinfo {author} {\bibfnamefont {D.}~\bibnamefont
  {Berardan}}, \bibinfo {author} {\bibfnamefont {A.}~\bibnamefont {Meena}},
  \bibinfo {author} {\bibfnamefont {S.}~\bibnamefont {Franger}}, \bibinfo
  {author} {\bibfnamefont {C.}~\bibnamefont {Herrero}}, \ and\ \bibinfo
  {author} {\bibfnamefont {N.}~\bibnamefont {Dragoe}},\ }\href {\doibase
  https://doi.org/10.1016/j.jallcom.2017.02.070} {\bibfield  {journal}
  {\bibinfo  {journal} {Journal of Alloys and Compounds}\ }\textbf {\bibinfo
  {volume} {704}},\ \bibinfo {pages} {693 } (\bibinfo {year}
  {2017})}\BibitemShut {NoStop}%
\bibitem [{\citenamefont {R\'{a}k}\ \emph {et~al.}(2018)\citenamefont
  {R\'{a}k}, \citenamefont {Maria},\ and\ \citenamefont
  {Brenner}}]{rak_2018_ml}%
  \BibitemOpen
  \bibfield  {author} {\bibinfo {author} {\bibfnamefont {Z.}~\bibnamefont
  {R\'{a}k}}, \bibinfo {author} {\bibfnamefont {J.-P.}\ \bibnamefont {Maria}},
  \ and\ \bibinfo {author} {\bibfnamefont {D.}~\bibnamefont {Brenner}},\ }\href
  {\doibase https://doi.org/10.1016/j.matlet.2018.01.111} {\bibfield  {journal}
  {\bibinfo  {journal} {Materials Letters}\ }\textbf {\bibinfo {volume}
  {217}},\ \bibinfo {pages} {300 } (\bibinfo {year} {2018})}\BibitemShut
  {NoStop}%
\bibitem [{\citenamefont {Fukushima}\ \emph {et~al.}(2017)\citenamefont
  {Fukushima}, \citenamefont {Katayama-Yoshida}, \citenamefont {Sato},
  \citenamefont {Ogura}, \citenamefont {Zeller},\ and\ \citenamefont
  {Dederichs}}]{fukushima_2017_jpsj}%
  \BibitemOpen
  \bibfield  {author} {\bibinfo {author} {\bibfnamefont {T.}~\bibnamefont
  {Fukushima}}, \bibinfo {author} {\bibfnamefont {H.}~\bibnamefont
  {Katayama-Yoshida}}, \bibinfo {author} {\bibfnamefont {K.}~\bibnamefont
  {Sato}}, \bibinfo {author} {\bibfnamefont {M.}~\bibnamefont {Ogura}},
  \bibinfo {author} {\bibfnamefont {R.}~\bibnamefont {Zeller}}, \ and\ \bibinfo
  {author} {\bibfnamefont {P.~H.}\ \bibnamefont {Dederichs}},\ }\href {\doibase
  10.7566/JPSJ.86.114704} {\bibfield  {journal} {\bibinfo  {journal} {Journal
  of the Physical Society of Japan}\ }\textbf {\bibinfo {volume} {86}},\
  \bibinfo {pages} {114704} (\bibinfo {year} {2017})}\BibitemShut {NoStop}%
\end{thebibliography}%

\end{document}